\documentclass[12pt, a4paper]{article}
\pdfoutput=1

\usepackage[height=24cm,width=17cm]{geometry}
\usepackage{amsfonts}
\usepackage{amsmath}
\usepackage{amssymb}
\usepackage{dcolumn}
\usepackage{bm,here}
\usepackage[sort&compress, numbers, merge]{natbib}
\usepackage{subfig}
\usepackage{comment}
\usepackage{ifpdf}
\usepackage{slashed}
\usepackage{colortbl}
\usepackage{xcolor}
\usepackage[mathscr]{eucal}
\usepackage[utopia]{mathdesign}

\ifpdf        
  \usepackage{graphicx}     
  \usepackage[bookmarksopen,colorlinks=true,linkcolor=bblue,
  citecolor=ppink,urlcolor=ppink,linktoc=all]{hyperref}
\else     
  \usepackage[dvipdfmx]{graphicx}     
  \usepackage[dvipdfmx,bookmarksopen,colorlinks=true,linkcolor=bblue,
  citecolor=ppink,urlcolor=ppink,linktoc=all]{hyperref}
\fi

\usepackage{multicol}
\definecolor{red}{rgb}{1,0,0}
\definecolor{blue}{rgb}{0,0,1}
\definecolor{orange}{rgb}{1,0.5,0}
\definecolor{ppink}{rgb}{1,0.4,0.4}
\definecolor{bblue}{rgb}{0.284602,0.317763,0.963947}
\usepackage{framed}
\definecolor{shadecolor}{rgb}{0.95,0.95,0.95}

\makeatletter
\@addtoreset{equation}{section}

\makeatother

\newcommand{\prn}[1]{\left( {#1} \right)}
\newcommand{\com}[1]{\left[ {#1} \right]}
\newcommand{\lmk}{\left(}  
\newcommand{\rmk}{\right)}
\newcommand{\lkk}{\left[}  
\newcommand{\rkk}{\right]}

\newcommand{\del}{\partial}

\newcommand{\der}{\partial}  
\newcommand{\dd}{\mathrm{d}}

\newcommand{\abs}[1]{\left\vert {#1} \right\vert}
\newcommand{\mphi}{m_\phi}

\newcommand{\Min}{{\rm Min}}

\newcommand{\cphi}{\varphi}

\def\eq#1{Eq.~(\ref{#1})}
\newcommand{\bea}{\begin{array}}
\newcommand{\eea}{\end{array}}
\newcommand{\beq}{\begin{eqnarray}}
\newcommand{\eeq}{\end{eqnarray}}

\newcommand{\rad}{\psi}

\begin{document}

\begin{titlepage}

\begin{flushright}
IPMU16-0197
\end{flushright}

\vskip 2.5cm
\begin{center}
{\Huge \bf 
On Longevity of I-ball/Oscillon 
}
\vskip .65in

{\large Kyohei Mukaida$^{a}$,
Masahiro Takimoto$^{b,c}$
and Masaki Yamada$^{d,e}$ 
}

\vskip .35in
\begin{tabular}{ll}
$^{a}$ &\!\! {\em Kavli IPMU (WPI), UTIAS,}\\
&\!{\em The University of Tokyo, Kashiwa, Chiba 277-8583, Japan}\\[.3em]
$^{b}$&\!\! {\em Theory Center, KEK, 1-1 Oho, Tsukuba, Ibaraki 305-0801, Japan}\\[.3em]
$^{c}$&\!\! {\em Department of Particle Physics and Astrophysics,}\\
&\!{\em Weizmann Institute of Science, Rehovot 7610001, Israel}\\[.3em]
$^{d}$ &\!\! {\em Institute of Cosmology, Department of Physics and Astronomy,}\\
&\!{\em Tufts University, Medford, MA  02155, USA}\\[.3em] 
$^{e}$ &\!\! {\em Department of Physics, Tohoku University, Sendai, Miyagi 980-8578, Japan}
\end{tabular}

\vskip .75in

\begin{abstract}  
\noindent
We study I-balls/oscillons, which are long-lived, quasi-periodic, and spatially localized solutions in real scalar field theories. Contrary to the case of Q-balls, there is no evident conserved charge that stabilizes the localized configuration. Nevertheless, in many classical numerical simulations, it has been shown that they are extremely long-lived. In this paper, we clarify the reason for the longevity, and show how the exponential separation of time scales emerges dynamically. Those solutions are time-periodic with a typical frequency of a mass scale of a scalar field. This observation implies that they can be understood by the effective theory after integrating out relativistic modes. We find that the resulting effective theory has an approximate global U$(1)$ symmetry reflecting an approximate number conservation in the non-relativistic regime. As a result, the profile of those solutions is obtained via the bounce method, just like Q-balls, as long as the breaking of the U$(1)$ symmetry is small enough. We then discuss the decay processes of the I-ball/oscillon by the breaking of the U$(1)$ symmetry, namely the production of relativistic modes via number violating processes.  We show that the imaginary part is exponentially suppressed, which explains the extraordinary longevity of I-ball/oscillon. In addition, we find that there are some attractor behaviors during the evolution of I-ball/oscillon that further enhance the lifetime. The validity of our effective theory is confirmed by classical numerical simulations. Our formalism may also be useful to study condensates of ultra light bosonic dark matter, such as fuzzy dark matter, and axion stars, for instance. 
\end{abstract}

\end{center}
\end{titlepage}

\tableofcontents
\thispagestyle{empty}
\newpage
\setcounter{page}{1}

\section{Introduction}
\label{sec:intro_sum}

Condensates of scalar fields play important roles in the early Universe. One of the most prominent examples is the inflaton field which causes the accelerated expansion of the Universe, \textit{i.e.}, inflation~\cite{Guth:1980zm, Linde:1981mu}, and may seed primordial density fluctuations~\cite{Ade:2013zuv}. A curvaton field~\cite{Enqvist:2001zp, Lyth:2001nq, Moroi:2001ct} is another candidate to generate the primordial density fluctuations. As the Higgs field in the Standard Model, some scalar field may realize a phase transition that leads to a spontaneous symmetry breaking (SSB). One of the most important examples of this kind is the SSB of Peccei-Quinn (PQ) symmetry, which is introduced to explain the strong CP problem~\cite{Peccei:1977hh}. The SSB results in a prediction of a pseudo-Nambu-Goldstone boson called axion~\cite{Weinberg:1977ma} and it is known that the sizable amount of axion can be produced in the form of condensate by the misalignment mechanism~\cite{Preskill:1982cy, Abbott:1982af, Dine:1982ah}. In the Affleck-Dine baryogenesis scenario~\cite{Affleck:1984fy, Murayama:1993em, Dine:1995kz}, baryonic U$(1)$ charged scalar condensates are indispensable to generate the baryon asymmetry of the Universe.

Some scalar fields come to form (quasi-)stable and localized objects in the early stage of the Universe. Since the formation and time evolution of such localized objects may significantly affect the cosmological scenarios, it is important to understand their dynamics. In general, their existence is ensured by some conserved quantities. For example, the conservation of topological numbers will lead to the formation of monopoles, cosmic strings or domain walls, that are called as topological defects~\cite{Zeldovich:1974uw}.  Such objects are particularly important to consider axion cosmology because cosmic strings form at the SSB of PQ symmetry and domain walls may form at the QCD phase transition depending on the model~\cite{Sikivie:1982qv, Vilenkin:1982ks}. Another example of localized objects is so-called Q-balls~\cite{Coleman:1985ki, Kusenko:1997si,Enqvist:1997si,Enqvist:1998en,Kasuya:1999wu,Kasuya:2000wx,Kasuya:2001hg} which appear in complex scalar theories and  whose existence is ensured by the global U$(1)$ symmetry.  In any case, the conserved quantity seems to be the fundamental essence of their stability.

I-balls/oscillons~\cite{Bogolyubsky:1976nx,Segur:1987mg,Gleiser:1993pt,Copeland:1995fq,Gleiser:1999tj, Honda:2001xg,Kasuya:2002zs,Gleiser:2004an,Fodor:2006zs,Hindmarsh:2006ur,Saffin:2006yk, Fodor:2008du,Gleiser:2008ty,Fodor:2008du,Fodor:2009kf,Gleiser:2009ys,Amin:2010jq,Amin:2011hj,Salmi:2012ta,Andersen:2012wg, Lozanov:2014zfa,Saffin:2014yka,Mukaida:2014oza,Kawasaki:2015vga}  are long-lived (but not absolutely stable), quasi-periodic, and spatially localized objects which appear in real scalar field theories if the potential is  shallower than quadratic. The existence of such objects seems to be mysterious because there are no evident conserved quantities related to them. Nevertheless, many classical numerical studies have revealed that  they are extremely long-lived. In Ref.~\cite{Kasuya:2002zs}, it was discussed that the existence of I-balls/oscillons may be related to the (approximate) conservation of the adiabatic charge $I$ in the non-relativistic regime. In Ref.~\cite{Mukaida:2014oza},  two of the authors found that the existence of  I-ball/oscillon may be understood by the approximate number conservation in non-relativistic regime~\cite{Berges:2014xea, Davidson:2014hfa, Braaten:2016kzc}. Those observations suggest that an approximately conserved quantity (number or adiabatic charge) in the non-relativistic regime of real scalar theories may play a major role in the existence and longevity of I-balls/oscillons.

The main purpose of this paper is to clarify the reason for the longevity of I-balls/oscillons, revealed in many classical numerical simulations, from the viewpoint in \cite{Mukaida:2014oza} (See Refs.~\cite{Segur:1987mg,Gleiser:2008ty,Fodor:2008du,Fodor:2009kf} for different approaches). To do so, we construct an effective theory of real scalar field theories with polynomial potentials suitable to describe the dynamics in the non-relativistic regime. By integrating out relativistic modes, we obtain a non-relativistic effective theory. The resulting effective theory comes to have an approximate global U$(1)$ symmetry, which reflects the approximate number conservation in the non-relativistic regime (See Ref.~\cite{Moore:2015adu} for the same viewpoint in turbulence). In this approach, I-ball/oscillon can be understood by the same way as that of Q-ball, as long as the breaking of the U(1) symmetry is small enough. This breaking is imprinted in the imaginary part of the effective theory, which reflects the production of relativistic particles via number violating processes. By omitting all the loop diagrams in the effective action, we explicitly show the smallness of the imaginary part. This explains the reason why the \textit{classical} decay rate is so suppressed. (See Refs.~\cite{Hertzberg:2010yz,Kawasaki:2013awa,Saffin:2016kof} and Sec.~\ref{sec:conclusion} for \textit{quantum} decays.) In particular, we find that there are some approximate attractor behaviors during the evolution of I-ball/oscillon. As a result, the lifetime of I-ball/oscillon is enhanced by many orders of magnitude.  We also confirm the validity of our description by comparing the results of our effective theory with the numerical simulations of the original relativistic scalar field theory.

This paper is organized as follows. In Sec.~\ref{sec:nr_cond}, we first explain how to construct the effective theory of the scalar field in the non-relativistic regime. Then, we demonstrate how to obtain the profile of I-balls/oscillons in the same way as Q-balls, and why they are so long-lived from the viewpoint of the approximate conservation of the U(1) symmetry.
In Sec.~\ref{sec:num_sim}, we compare the results of our effective theory obtained in Sec.~\ref{sec:nr_cond} with those of numerical simulations of the original relativistic scalar field theory. Sec.~\ref{sec:conclusion} is devoted to conclusions and discussion. Possible impacts of \textit{quantum} decays are briefly mentioned.

\section{I-ball/Oscillon as Non-relativistic Condensates}
\label{sec:nr_cond}

In this section, we will argue that I-balls/oscillons can be regarded as \textit{pseudo} nontopological solitons in the non-relativistic limit of real scalar field theories. This is because an approximate U(1) symmetry associated with the number conservation emerges in the non-relativistic regime. Since this symmetry is not exact, those \textit{pseudo} nontopological solitons eventually decay, whose time scale is controlled by the breaking of this symmetry.

First, we will sketch the way how to derive a non-relativistic effective theory from a relativistic real scalar. After integrating out relativistic modes, we will find that the approximate U(1) symmetry appears with respect to the number conservation. The breaking of U(1) symmetry, which is inevitable because the interacting relativistic real scalar field does not conserve its number, can be identified as the imaginary part of the effective action. The breaking corresponds to the production of relativistic modes.

Next, we discuss the property of I-balls/oscillons, treating the imaginary part perturbatively. Namely, we first find a non-trivial spatially localized solution, \textit{i.e.,}~I-ball/oscillon, neglecting the imaginary part, and then discuss its decay taking account of the imaginary part at the first order. If the decay time is much longer than the frequency of these solutions, we can regard them as \textit{pseudo} nontopological solitons associated with the approximate U(1) symmetry. Interestingly, the properties of I-balls/oscillons strongly depend on whether or not the scalar potential respects a $\mathbb{Z}_2$ symmetry at the vacuum, $(\phi - \langle \phi \rangle) \mapsto - (\phi - \langle \phi \rangle)$.\footnote{
	The potential with an explicit $\mathbb{Z}_2$ breaking term falls into this case. 	In addition, even if the potential respects the $\mathbb{Z}_2$ symmetry, the vacuum state can break it spontaneously. This case also falls into the $\mathbb{Z}_2$ breaking one in our classification.
}

Finally, we discuss the \textit{classical} decay of I-ball/oscillon via the breaking of U$(1)$ symmetry by employing a simple model. We explicitly show the smallness of the imaginary part, which indicates the reason why the I-balls/oscillons are so long-lived in many classical numerical simulations. We also find several critical values of the charge $Q$ where the dominant decay channel is suppressed. As a result, I-balls/oscillons stay at such critical points in most of their lifetime. Those results are confirmed by numerical simulations of the original relativistic scalar field theory in Sec.~\ref{sec:num_sim}.

\subsection{Non-relativistic effective field theory}
\label{sec:non_rela_eft}

Let us start with the following relativistic real scalar field theory:
\begin{align}
	\mathcal L = \frac{1}{2} \der_\mu \phi \der^\mu \phi - \frac{1}{2} m_\phi^2 \phi^2 - V_\text{int} (\phi),
	\label{lagrangian}
\end{align}
where $\phi$ is a real scalar field, $m_\phi$ is its mass, and $ V_\text{int}$ is the interaction term. The goal of this subsection is to derive an effective action which describes the non-relativistic motion of $\phi$ by integrating out relativistic modes from this Lagrangian.

We divide the scalar field $ \phi$ into two parts; the non-relativistic part and the other:
\begin{align}
	\phi (x) = \phi_\text{NR} (x) + \delta \phi (x),
	\label{eq:nr_rela}
\end{align}
where
\begin{align}
	\phi_\text{NR} (x) \equiv \int_{K \in \,\text{NR}} e^{- i K \cdot x} \phi (K),
	\quad
	\delta \phi (x) \equiv \int_{K \in \,\overline{\text{NR}}} e^{- i K \cdot x} \phi (K),
\end{align}
with $(K^\mu)=(k^0, \bm{k})$. $\phi (K)$ is the Fourier coefficient of the real scalar field, which satisfies $\phi (K) = \phi^\ast (-K)$. The word ``NR'' indicates the region which is close to the on-shell pole of non-relativistic excitations, \textit{i.e.,} $\text{NR} \equiv \{ (k_0 ,\bm{k})\vert \, \pm k_0 \sim m_\phi + \mathcal O (m_\phi v^2), \, \pm \bm{k} \sim \mathcal O (m_\phi \bm{v}) \}$ with $v \equiv \abs{\bm{v}} \ll 1$. $\overline{\text{NR}}$ is defined as its complementary set. For later convenience, we further split the non-relativistic part into two parts:
\begin{align}
	\phi_\text{NR} (x) = \frac{1}{2} \com{ \Psi (t, \bm{x}) e^{-i m_\phi t} + \text{H.c.}},
\end{align}
where
\begin{align}
	\Psi (t,\bm{x}) = \int_{K \in \, \text{NR}_+} e^{ - i (k_0 - m_\phi)t + i \bm{k} \cdot \bm{x}} \phi (K).
	\label{eq:nr}
\end{align}
Here $\text{NR}_+$ represents a non-relativistic region with positive energy $k_0 > 0$. By definition, $\Psi$ is a slowly varying field
\begin{align}
	\abs{ \frac{\der^2 \Psi}{ \der (m_\phi t )^2}}
	\ll 
	\abs{ \frac{\der\Psi}{ \der (m_\phi t )}} 
	\ll 
	\abs{\Psi},
\end{align}
and also its spatial gradient is small
\begin{align}
	\abs{ \frac{\nabla^2 \Psi}{\Psi} }
	\ll
	m_\phi^2.
\end{align}
The initial condition of the original  real scalar field theory is mapped by
\begin{align}
	\left. \phi \right|_\text{ini} \simeq \Re \com{\left.\Psi \right|_\text{ini}}, \quad
	\left. \dot \phi \right|_\text{ini} \simeq m_\phi \Im \com{ \left. \Psi \right|_\text{ini} }.
	\label{eq:initial cond}
\end{align}

We can derive the non-relativistic effective field theory by integrating out $\delta \phi$. Throughout this paper, we focus on the classical longevity of I-ball/oscillon in order to clarify the reason why the I-balls/oscillons are so long-lived in many classical numerical studies. Thus, we omit the loop contributions from the effective action so that we can obtain the \textit{classical} non-relativistic effective field theory. Possible quantum effects will be discussed in Sec.~\ref{sec:conclusion}. The property of the effective field theory depends on whether or not it respects a $\mathbb{Z}_2$ symmetry at the vacuum, \textit{i.e.,} $(\phi - \langle \phi \rangle ) \mapsto - (\phi - \langle\phi \rangle)$. We illustrate what we are going to work on by taking simple examples for each case.

\subsubsection*{With $\mathbb{Z}_2$ symmetry}

First, let us assume that the theory respects a $\mathbb{Z}_2$ symmetry at the vacuum: $(\phi - \langle \phi \rangle) \mapsto - (\phi - \langle \phi \rangle)$. Without loss of generality, we take $\langle \phi \rangle = 0$. To be concrete, suppose that $V_\text{int} = - g_4 \phi^4 / 4 + \cdots$ dominates the potential. Here note that higher order terms are implicit, which are required to stabilize the vacuum at $\langle \phi \rangle = 0$. The relevant interaction terms are
\begin{align}
\mbox{
	\includegraphics[width=\textwidth]{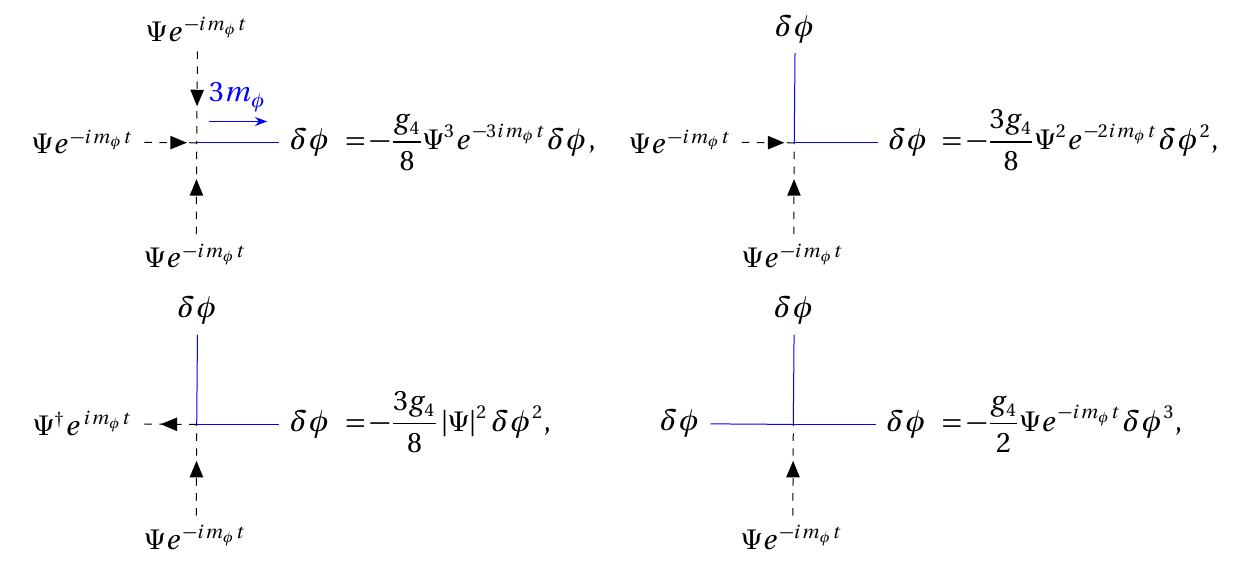}
}
\end{align}
and those conjugates.\footnote{
	The Hermite conjugate of $(3 g_4 / 8) \abs{\Psi}^2\delta \phi^2$ should also be included.
}
Here a dashed line with an arrow represents a non-relativistic mode $ \Psi$ or $\Psi^\dag$,
and a blue line is the other $\delta \phi$.
Note that the terms, like $\delta \phi |\Psi|^{2n}$, are forbidden by the energy conservation
because the operator $|\Psi|^{2n}$ cannot carry relativistic energies.

To obtain an effective theory which describes the dynamics of non-relativistic modes,
we only keep $\Psi e^{- i m_\phi t}$ and $\Psi^\dag e^{i m_\phi t}$ 
as external lines and integrate out $\delta \phi$.
After integration, we obtain vertices such as $\Psi^n e^{- n i m_\phi t} \Psi^{\dag^m} e^{m i m_\phi t}$.
Since the non-relativistic fields, $\Psi$ ($\Psi^\dag$), cannot provide incoming (outgoing) energies 
comparable to the mass scale $m_\phi$, vertices except $ n = m$ vanish owing to the energy conservation.
As a result, all the interaction terms involve a pair of $\Psi \Psi^\dag$,
which respects the U(1) symmetry: $\Psi \mapsto e^{i \theta} \Psi$.
This U(1) symmetry is nothing but the number conservation of non-relativistic particles.

To see how to integrate out $\delta \phi$, 
let us study the following diagram:\footnote{
	Here we symbolically write the propagator as $-1/(\Box + m_\phi^2)$,
	but note that the $i \varepsilon$-term should be included properly,
	which reflects the physical boundary condition of this setup.
	See also Sec.~\ref{sec:decay}.
}
\begin{align}
\mbox{
	\includegraphics[width=0.7\textwidth]{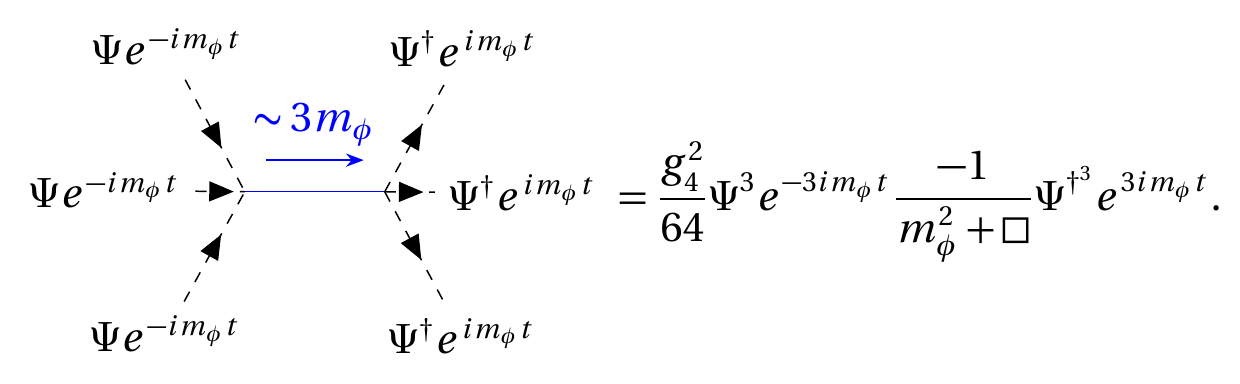}
}
	\label{eq:3to1}
\end{align}
Let us first discard terms where $\der_\mu$ acts on $\Psi$ because it is slowly varying and has small gradients.
The leading order real part contribution comes from the term where $\der_t^2$ acts on $e^{3 i m_\phi t}$.
One finds the following corrections to the effective action,
$\delta \mathcal L \supset - g_4^2 \abs{\Psi}^6 / (512 m_\phi^2) + \cdots $,
where ellipses represent higher order contributions.

However, the number conservation is not exact for an interacting relativistic real scalar.
The breaking of U(1) symmetry is imprinted in the imaginary part of the effective action after integration.
Such an imaginary part comes from cuttings of diagrams which stand for the production of relativistic modes.
To obtain the leading order imaginary part from the diagram of \eq{eq:3to1},
we have to keep terms where $\der_\mu$ act on $\Psi$,
contrary to the real part contributions.
Although its fraction is tiny, $\Psi$ may contain higher momentum modes.
If this is the case, the pole of Eq.~\eqref{eq:3to1} can yield the imaginary part of the effective action,
which indicates the production of relativistic modes with $k_0 \sim 3 m_\phi, k \equiv \abs{\bm{k}} \sim 2 \sqrt{2} m_\phi$.

In the following, we first neglect the imaginary part and find a spatially localized energetically favored state
in Sec.~\ref{sec:iball_prop}.
Then, we discuss its decay.
Treating the classical lump as a background, we will see that a tiny fraction of the obtained solution
can hit the pole and produces relativistic modes in Sec.~\ref{sec:decay}.

\subsubsection*{Without $\mathbb{Z}_2$ symmetry}
Then, let us add terms which break the $\mathbb{Z}_2$ symmetry at the vacuum: 
$(\phi - \langle \phi \rangle) \mapsto - (\phi -  \langle \phi \rangle)$.
Again we shift the field so that $\langle \phi \rangle = 0$ without loss of generality.
In this case, we have interaction terms, like $\abs{\Psi}^n \delta \phi$, which do not contain a
rapidly oscillating term, $e^{n i m_\phi t}$.
The term, $\delta \phi (t, \bm{x})$, is not relativistic rather varying slowly.
Nevertheless, it is far below the pole of non-relativistic excitations
because its typical energy is $|k_0| \sim m_\phi v^2 \ll m_\phi$.
We denote such contributions as $\delta \phi_V$ so as to distinguish them
from relativistic fluctuations in the following.

To be concrete, suppose that we have a cubic interaction, $g_3 \phi^3 / 3$.
It yields the following interaction term involving $\delta \phi_V$:
\begin{align}
	\mbox{\includegraphics[width=0.42\textwidth]{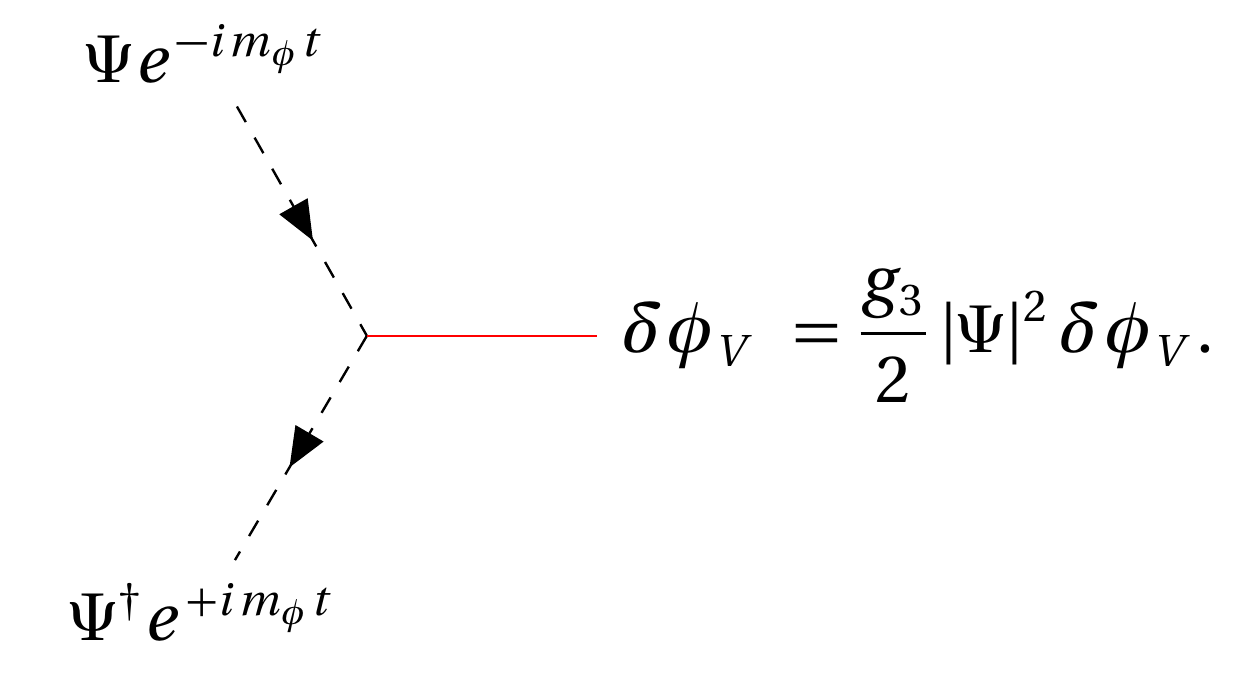}}
\end{align}
Here the red line is $\delta \phi_V$, which is slowly varying but far below the pole of
non-relativistic excitations.
Integrating out such fluctuations,
one finds the following diagram
\begin{align}
	\mbox{\includegraphics[width=0.55\textwidth]{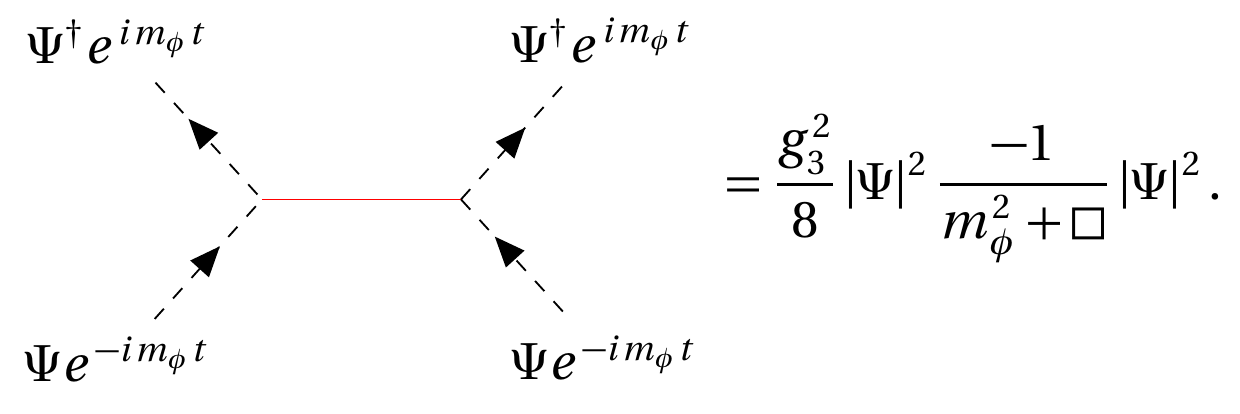}}
		\label{eq:2to1to2}
\end{align}
The additional $1/2$ comes from the symmetry factor of this diagram.
Here, we can neglect the derivatives in the denominator 
in the leading approximation. 
As an another approach, it is instructive to see the equation of motion for $\delta \phi_V$.
Suppose that the system somehow becomes (quasi-)static, which is the case for the I-ball and oscillon solution.
Then, $\delta \phi_V$ should satisfy the following equation:
\begin{align}
	\delta \phi_V = - \frac{g_3 \abs{\Psi}^2}{2 m_\phi^2} + \cdots.
	\label{eq:delta phi_V}
\end{align}
Substituting this into the Lagrangian, we obtain the same result as \eq{eq:2to1to2}. 
One can see that $\phi$ gets a constant shift due to the condensate of the non-relativistic field, $\Psi$.
Note that its amplitude is smaller than $\Psi$, \textit{i.e.,} $|\delta \phi_V| \ll |\Psi|$, if the coupling is small.

We also have the following diagram involving an intermediate relativistic fluctuation,
\begin{align}
	\mbox{\includegraphics[width=0.67\textwidth]{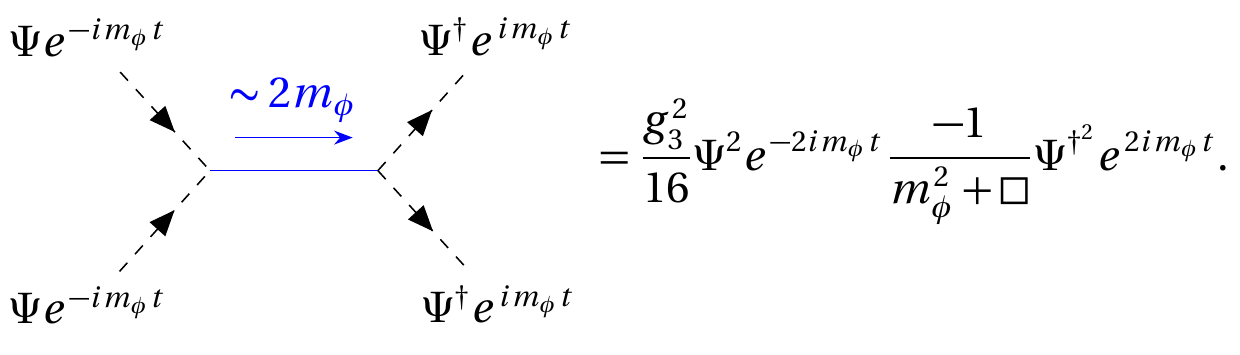}}
		\label{eq:2to1}
\end{align}
Again, one can obtain the leading correction to the real part of the effective action,
omitting terms $\der_\mu$ acting on $\Psi$.
We find the following correction to the effective action:
$\delta \mathcal L \supset 5 g_3^2/ (48 m_\phi^2) \abs{\Psi}^4 + \cdots$.
After integration, one can see that the vertices only involve pairs of $\Psi \Psi^\dag$
owing to the energy conservation.
Therefore, the real part of the effective action respects the U(1) symmetry:
$\Psi \mapsto e^{i \theta} \Psi$.

On the other hand, the leading contribution to the imaginary part comes from
the cutting of Eq.~\eqref{eq:2to1}.\footnote{
	It may be unlikely to hit the pole of Eq.~\eqref{eq:2to1to2}
	because not only the spatial gradients but the incoming (outgoing) energy is
	expected to be	much smaller than the mass scale $m_\phi$.
}
Plugging a spatially localized solution of $\Psi$ 
which is obtained by neglecting the imaginary part,
we will see that the tiny fraction contains high momentum modes with $p \sim \sqrt{3} m_\phi$.
Hence, the production of relativistic particles with $p_0 \sim 2 m_\phi$ and $p \sim \sqrt{3} m_\phi$
takes place. See Sec.~\ref{sec:decay} for details.

\subsubsection*{Summary}

Here we summarize basic properties of the non-relativistic effective field theory
which is obtained from integrating out $\delta \phi$ and also $\delta \phi_V$
from a relativistic real scalar field theory.
As already explained, all the vertices appear with pairs of $\Psi \Psi^\dag$ owing to the energy conservation,
and hence the effective theory respects the U(1) symmetry: $\Psi \mapsto e^{i \theta}\Psi$.
This symmetry corresponds to the number conservation of non-relativistic particles.
Since an interacting relativistic scalar does not conserve the particle number,
this symmetry should not be exact.
The breaking of U(1) is imprinted in the imaginary part of the effective action,
which stands for the production of relativistic modes.

To sum up, the effective theory for non-relativistic fields, $\Psi$, takes the form:
\begin{align}
	S_\text{NR} [\Psi] = \int_x \frac{1}{4} \com{ \Psi^\dag \prn{ 2 i m_\phi \der_t - \der_t^2 + \nabla^2 }\Psi 
	- V_\text{eff} (\abs{\Psi})} - i \Gamma[\Psi].
	\label{eq:S_NR}
\end{align}
We adopt the following potential as an example in the following discussion:
\begin{align}
	V (\phi) = \frac{g_3}{3} \phi^3 - \frac{g_4}{4} \phi^4 + \frac{g_6}{6} \phi^6.
	\label{V_phi}
\end{align}
By performing the same procedure demonstrated around Eqs.~\eqref{eq:3to1}, \eqref{eq:2to1to2} and \eqref{eq:2to1},
one may obtain the effective potential perturbatively after some algebras:
\begin{align}
	V_\text{eff} (\abs{\Psi}) = 
	\prn{-\frac{5g_3^2}{12m_\phi^2} - \frac{3g_4}{8} } \abs{\Psi}^4 
	+ \prn{
	\frac{5g_6}{24} 
	+\frac{g_4^2}{128m_\phi^2}
	} 
	\abs{\Psi}^6 + \cdots.
	\label{V_eff}
\end{align}
Note that a factor of $-1/4$ should be multiplied to get the effective potential, $V_\text{eff}$, from the corrections to
the Lagrangian $\delta \mathcal L$, since we have defined the effective action by Eq.~\eqref{eq:S_NR}.
We have omitted higher order terms;
such as  
those with higher order in the coupling or
those with $\der_\mu$ acting on $\Psi$,
so as to get the leading order corrections to the effective potential,
for $\Psi$ varies slowly and has small spatial gradients.\footnote{
	Then, the resultant effective potential is the same as one obtained from the so-called $\epsilon$-expansion. Thus, the $\epsilon$-expansion can be regarded as a perturbative computation of the effective potential in the non-relativistic effective field theory, where derivatives acting on $\Psi$ are neglected. To make the comparison easier, we do not rescale $\Psi$ to make its kinetic term canonical. 
}
To discuss higher order terms in the coupling consistently,
terms with $\der_\mu$ acting on $\Psi$
should also be included.
See App.~\ref{sec:higher order} for details.

The imaginary part of the effective action, $\Gamma [\Psi]$, may be dominated by
the cutting diagrams of Eqs.~\eqref{eq:3to1} or \eqref{eq:2to1}
for \textit{classical} decays, as discussed in Sec.~\ref{sec:decay}.

\subsection{I-ball/Oscillon as a \textit{pseudo} nontopological soliton}
\label{sec:iball_prop}

There exist two conserved quantities, its energy and charge associated with the U(1) symmetry, 
in the effective theory, if we neglect the production of relativistic modes induced via the imaginary part.
The goal of this subsection is to derive a non-trivial spatially localized solution 
as an energetically favored state with a fixed charge.
This treatment is justified a posteriori in Sec.~\ref{sec:decay},
where we show that the decay rate induced by this imaginary part
is much slower than the typical frequency of these solutions.
Thus, at each time step, we can safely regard its energy and charge as conserved quantities 
and discuss its decay as an adiabatic process.

\subsubsection*{Conserved Quantities}

The U(1) charge associated with $\Psi \mapsto e^{i \theta} \Psi$ is given by
\begin{align}	
	Q = \frac{1}{4} 
	i \int_{\bm{x}} \com{\Psi^\dag \prn{ {\der}_t  \Psi - im_\phi \Psi } - \Psi \prn{ {\der}_t  \Psi^\dag + im_\phi \Psi^\dag } }, 
\end{align}
where $\int_{\bm{x}} \equiv \int \dd^3 \bm{x}$.
And the energy is
\begin{align}
	E = 
	\frac{1}{4}
	\int_{\bm{x}} \com{ 
		\abs{ \der_t \Psi - im_\phi \Psi }^2  +
		\der_{\bm{x}} \Psi^\dag \der_{\bm{x}} \Psi + m_\phi^2 \abs{\Psi}^2 + V_\text{eff} (\abs{\Psi})
	}.
	\label{eq:energy}
\end{align}
The initial condition given in Eq.~\eqref{eq:initial cond} implies that
this system is necessarily accompanied by a non-vanishing charge $Q$.
Note that the factor $1/4$ should be included because we use the non-canonical kinetic term in Eq.~(\ref{eq:S_NR}).

\subsubsection*{I-ball/Oscillon as a projection of Q-ball}
Let us seek the energetically favored configuration under a fixed charge $Q$,
as in the case of Q-ball.
Introducing a Lagrange multiplier $\omega$, all one has to do is to find the solution 
which minimizes~\cite{Coleman:1985ki,Kusenko:1997ad,Enqvist:2003zb} 
\begin{align}
	I &=E + \omega \prn{Q - \frac{1}{4} 
	i \int_{\bm{x}}  \com{\Psi^\dag \prn{ {\der}_t  \Psi - im_\phi \Psi } - \Psi \prn{ {\der}_t  \Psi^\dag + im_\phi \Psi^\dag } } } \nonumber\\
	&= 
	\frac{1}{4}
	\int_{\bm{x}} \com{
		\abs{ \der_t \Psi - i \prn{m_\phi - \omega} \Psi }^2 + \der_{\bm{x}} \Psi^\dag \der_{\bm{x}} \Psi  
		+ \prn{ m_\phi^2 - \omega^2} \abs{\Psi}^2 + V_\text{eff} (\abs{\Psi}) 
	} + \omega Q.
	\label{eq:gf}
\end{align}
The minimization of the first term yields $\Psi (t, \bm{x}) = e^{i \mu t} \rad (\bm{x}) /\sqrt{2}$ with $\mu \equiv m_\phi - \omega$.
Without loss of generality, one can take $\rad (\bm{x})$ as real.
Here note that the non-relativistic condition implies $\mu \ll m_\phi$.
Since the pressure costs energy, let us assume a spherically symmetric solution~\cite{Coleman:1977th}.
Then, varying Eq.~\eqref{eq:gf} with respect to $\rad (r)$,
one obtains the bounce equation:
\begin{align}
	0 = \com{ \frac{\der^2}{\der r^2} + \frac{2}{r} \frac{\der}{\der r} } \rad (r) 
	- \prn{ 2 \mu m_\phi  - \mu^2 } \rad (r) - \frac{\del V_\text{eff} (\rad)}{\del \rad}.
	\label{eq:bounce}
\end{align}
From the finiteness of energy, the following boundary conditions are imposed:
\begin{align}
	\lim_{r \to 0 } \frac{\del \rad (r)}{\del r} = \lim_{r \to \infty} \rad(r) = 0.
	\label{eq:bdry}
\end{align}
It can be regarded as an equation of motion for  a one-dimensional system with a time variable $r$ and a friction $2/ r$.
Therefore, a non-trivial solution may exist if 
the curvature of the ``effective'' potential at the origin, $- ( 2\mu m_\phi - \mu^2)$, 
is negative and the ``effective'' potential, 
$- (\mu m_\phi - \mu^2 /2) \rad^2 - V_\text{eff}$, becomes positive at $\rad \neq 0$.
If this condition is fulfilled, the bounce solution may exist, where
$\rad (\neq 0)$ starts to roll down its ``effective'' potential
while the friction $2/r$ dissipates its energy,
and it stops at $\rad = 0$ eventually.
This condition can be expressed as
\begin{align}
	0 < \mu m_\phi - \frac{\mu^2}{2} < - \Min \com{ \frac{V_\text{eff} (\rad) }{\rad^2} }.
	\label{eq:cond}
\end{align}
Once we find a family of solution for each $\mu$ as $\rad (r; \mu)$,
its charge and energy are obtained from
\begin{align}
	Q (\mu) & = \frac{1}{4}\omega \int 4 \pi r^2 \dd r\, \rad^2 (r; \mu), 
	\label{eq:charge}\\
	E (\mu) &= \frac{1}{4}
	\int 4\pi r^2 \dd r \com{ \frac{1}{2} \omega^2 \rad^2 (r; \mu) 
	+ \frac{1}{2} \prn{ \frac{\der }{\der r} \rad (r; \mu) }^2 + V_\text{eff} (\rad (r; \mu))}.
	\label{eq:qande_qball}
\end{align}
Here note again that  $\omega = m_\phi - \mu$.
The solution, $e^{- i \omega t} \rad (r; \mu)$, is nothing but the Q-ball solution,
which is a nontopological soliton in the presence of a conserved charge.
In particular, it has been shown that~\cite{Gulamov:2013cra}
\begin{align}
	\frac{\del E}{\del Q} = \omega \equiv m_\phi - \mu. 
	\label{eq:omega}
\end{align}

Finally, let us relate this solution to the original scalar field $\phi$.
Plugging this solution back to the definition given in Eqs.~\eqref{eq:nr_rela} and \eqref{eq:nr},
and omitting rapidly oscillating terms in $\delta \phi$,
we obtain the I-ball/oscillon solution as a projection of the Q-ball solution:
\begin{align}
	\phi (t,r ; \mu) = \Re \com{ \frac{1}{\sqrt{2}} e^{- i \prn{m_\phi - \mu} t} \rad (r; \mu) } + \delta \phi_V [\rad (r; \mu)],
\end{align}
where the shift is determined perturbatively [see \eq{eq:delta phi_V}] 
\begin{align}
	\delta \phi_V [\rad (r; \mu)]  = - \frac{g_3}{2 m_\phi^2} 
	\frac{\rad (r; \mu)^2}{2} + \cdots.
\end{align}
Therefore, 
the I-ball/oscillon can be regarded as a \textit{pseudo} nontopological soliton 
associated with the approximate U(1) symmetry in the non-relativistic regime.
In this sense, the I-ball/oscillon can be seen as a projection of the Q-ball.

\subsubsection*{Phase diagram and classical stability of Q-ball}

Here we give a numerical example to clarify the properties of I-ball/oscillon explained so far.
We assume $g_3 = 0$ in \eq{V_phi} for simplicity. 
The configuration of $\rad(r; \mu)$ can be calculated by solving 
\eq{eq:bounce} for given values of $\mu$, 
where $V_{\rm eff}$ is given by \eq{V_eff}. 
Then we can calculate the charge and energy of the configuration 
by Eqs.~(\ref{eq:charge}) and (\ref{eq:qande_qball}), respectively. 
We also define a typical radius of the configuration by 
$\rad (r = R_Q) = \rad(r=0) / e$. 
All quantities are functions of $\mu$, so that we can rewrite 
the amplitude of the configuration $\rad (r = 0)$ ($\equiv \rad_0$) and the radius $R_Q$ 
as a function of charge $Q$.

\begin{figure}[t]
\centering 
 \includegraphics[width=.60\textwidth ]{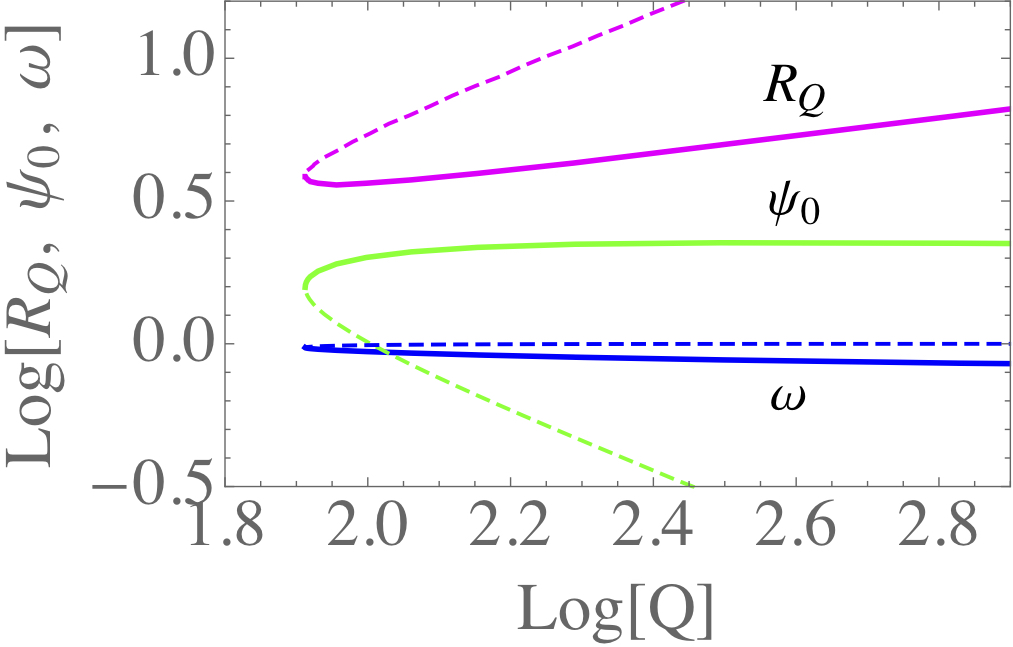} 
\caption{\small
Radius $R_Q$ (magenta line), amplitude $\rad_0$ (green line), and parameter $\omega$ (blue line) of I-ball/oscillon as a function of charge $Q$. 
We assume $g_4 = 1$ and $g_6 = 0.4$. 
There are two solutions for a given charge when it is larger than a critical value $Q_{\rm cr} \simeq 10^{1.91}$. 
Solutions are stable against a small perturbation 
on one branch with thick lines, 
while they are unstable on the other branch with dashed lines. 
}
  \label{fig:phase diagram}
\end{figure}

Figure~\ref{fig:phase diagram} shows the phase diagram of the I-ball/oscillon in this theory, 
where we assume $g_3 = 1$ and $g_6 = 0.4$. 
We find that 
there is no solution of \eq{eq:bounce} 
when its charge is less than a certain critical value $Q_{\rm cr} \simeq 10^{1.91}$. 
On the other hand, 
there are two solutions for a given charge $Q$ for $Q > Q_{\rm cr}$. 
On one branch, which is shown as thick lines in the figure, 
the amplitude of configuration and the radius increase as the charge increases, 
while on the other branch, which is shown as dashed lines in the figure, 
the amplitude decreases and the radius increases as the charge increases. 
The solutions on the former branch are stable against a small perturbation 
while the ones on the latter branch are unstable~\cite{Friedberg:1976me,Lee:1991ax,Tsumagari:2009zp,Enqvist:2003zb,Hertzberg:2010yz}.
Therefore, 
we have only one stable solution for a given $Q$ ($> Q_{\rm cr})$. 
When the theory has a charge-violating interaction, as we discuss in the next subsection, 
the configuration evolves along with the stable branch as its charge decreases. 
Note that there is no solution for $Q < Q_{\rm cr}$, 
which implies that the I-ball/oscillon disappears when its charge decreases to 
the critical value~\cite{Gleiser:2008ty,Amin:2010jq}.

The unstable solutions may be understood physically in the following way.  Suppose that the initial configuration is the corresponding Q-ball solution with a charge Q, whose energy is $E = \int^Q \dd Q' \omega (Q')$ [see \eq{eq:omega}]. Recall here that the Q-ball is characterized by a total charge $Q$ and its frequency $\omega (Q)$.
Then, let us perturb the solution with $\delta Q$.
In order for the Q-ball solution to be stable against the perturbation,
the energy of a Q-ball with charge $Q + \delta Q$ should be smaller than 
that of a Q-ball with charge $Q$ plus $\delta Q$ particles on it.
Therefore, the following inequality should hold for the stability of Q-balls
\begin{align}
	\int^{Q + \delta Q} \dd Q' \omega (Q') 
	<  \int^{Q} \dd Q' \omega (Q') 
	+ \delta Q \omega (Q)
	\quad \leftrightarrow \quad
	\frac{\der \omega}{\der Q} < 0.
	\label{eq:collapse}
\end{align}
As can be seen from a phase diagram in Fig.~\ref{fig:phase diagram},
the branch with thick lines respects this inequality, while the other does not.
In the next section, we will see that configurations close to the branch 
satisfying Eq.~\eqref{eq:collapse} are actually long-lived.
We also check numerically that those close to the other branch that violates Eq.~\eqref{eq:collapse}
decays immediately.

\subsection{Decay of I-ball/Oscillon via U(1) breaking} \label{sec:decay}

Now we are in a position to discuss the decay of I-ball/oscillon.
In the previous section, we have shown that there is a spatially localized energetically favored solution
by neglecting the U(1) breaking terms.
First, let us discuss its typical property, and see that it necessarily contains
a small amount of higher momentum modes.

To be specific, we characterize the bounce solution as follows:
\begin{align}
	\rad (r) \simeq \rad_0 e^{-r^2 / R^2},
	\label{eq:gaussian}
\end{align}
where $\rad_0$ is the amplitude at the center,
and $R$ is a typical size of the bounce solution.
The non-relativistic condition indicates that $m_\phi R \gg 1$.
Thus, the momentum of the I-ball/oscillon solution typically spreads over $k \lesssim 1/R \ll m_\phi$,
while its energy is concentrated on $k_0 = \omega (= m_\phi - \mu)$ with $\mu \sim 1/R \ll m_\phi$ as a delta function.
The crucial observation is that it also contains a tiny amount of high momentum modes with $k \sim m_\phi$,
nevertheless its energy is \textit{non-relativistic} $k_0 = m_\phi - \mu \simeq m_\phi$.
The attractive effective potential enables these apparently ``off-shell'' modes to exist  inside the I-ball/oscillon solution.

The existence of these modes alone does not spoil the stability of I-ball/oscillon,
since they cannot propagate outside the I-ball/oscillon as free particles
owing to the smallness of their energies~\cite{Hertzberg:2010yz}.
However, \textit{in combination with} the U(1) breaking terms, 
they play essential roles in production of relativistic modes.
This is because we can hit the poles of relativistic modes such as Eqs.~\eqref{eq:3to1} and \eqref{eq:2to1}
with the help of these high momentum modes.
We refer this process as \textit{decay via spatial gradients},
and discuss its nature by taking a simple example in the next subsection.
Through this process, the I-ball/oscillon gradually reduces its energy/number,
and eventually it exhibits sudden decay at a critically value of the charge, $Q_\text{cr}$,
below which there are no I-ball/oscillon solutions.
See a numerical justification of this picture in Sec.~\ref{sec:num_sim}.
It is noticeable that the energy spectrum of the I-ball/oscillon acquires a small width
because the delta function in energy is broadened by the decay processes via U(1) breaking terms.
Similar decay processes are discussed in the context of the axion star~\cite{Eby:2015hyx}.

\subsubsection*{Decay via spatial gradients}
As explained above,
the I-ball/oscillon solution contains a tiny amount of high momentum modes 
though its energy is still non-relativistic.
Performing the Fourier transform of Eq.~\eqref{eq:gaussian},
one can explicitly see this property:
\begin{align}
	\rad (\bm{k}) &= \int_{\bm{x}} e^{- i \bm{k} \cdot \bm{x}} \rad (\bm{x}) 
	\\
	&\simeq \lmk \frac{\pi}{2} \rmk^{3/2} R^3\rad_0 e^{- k^2 R^2 /4}, 
\end{align}
where we assume \eq{eq:gaussian} in the second line. 
There exist high momentum modes, $k \gtrsim m_\phi$, as a tail of the Gaussian distribution.

In the following, let us discuss how these high momentum modes induce the decay.
To be concrete, we consider the following interaction term as an example:
\begin{align}
	\mathcal L_\text{int} = \frac{g_{n+1}}{n+1} \phi^{n+1}.
	\label{eq:int_ex_or}
\end{align}
After splitting the scalar field into non-relativistic modes and the others,
one finds the following term which contains a linear term of $\delta \phi$:
\begin{align}
	\mathcal L_\text{int} 
	&\supset \frac{g_{n+1}}{2^{3n/2}} \rad^{n} e^{- i n m_\phi t} \delta \phi + \text{H.c.} 
	\\
	&\equiv J(x) \delta \phi + \text{H.c.}, 
	\label{eq:int_ex}
\end{align}
which plays important roles in the \textit{classical} decay.

Suppose that the I-ball/oscillon sits at the origin initially, and see how relativistic modes are produced in the presence of the I-ball/oscillon background. 
Let us emphasize that $\rad (r; \mu)$ is now time-dependent 
because of the decay process. 
To estimate the decay rate in such a situation, 
first note that the imaginary part of effective action [see \eq{eq:S_NR}] 
contributes to the time-dependence of the amplitude $\rad (r,t; \mu)$: 
\begin{align}
	&\frac{1}{4} \lkk \lmk 2 m_\phi i \frac{\del}{\del t} + \frac{\del^2}{\del t^2} + \nabla^2 \rmk \Psi (t, x) 
	- \frac{\del V_{\rm eff}}{\del \Psi^\dagger} \rkk - i \frac{\del \Gamma}{\del \Psi^\dagger} = 0 
	\\
	&\leftrightarrow 
	\frac{1}{2} m_\phi i \frac{\del}{\del t} \rad (r,t; \mu) - i \frac{\del \Gamma}{\del \rad} = 0, 
\end{align}
where we use the equation of motion of the real part in the second line 
and neglect a term proportional to $\del^2 \psi / \del t^2$ because the decay process is assumed to be much slower than the oscillation time scale $m_\phi$.\footnote{
	And in fact, we will see soon that the decay rate is much smaller than $m_\phi$ a posteriori.
}
Recalling the definition of the charge given in Eq.~\eqref{eq:charge},
one finds its time derivative as follows:
\begin{align}
	\dot Q 
	= \frac{1}{2} \omega \int_{\bm x} \rad \dot \rad 
	\simeq + \int 4 \pi r^2 \dd r\, \rad  
	\frac{\der \Gamma}{\der \rad}.
\end{align}
The imaginary part of effective action $i \Gamma$ comes from 
diagrams like Eqs.~\eqref{eq:3to1} and \eqref{eq:2to1}. 
For the interaction of \eq{eq:int_ex}, 
it is given by 
\begin{align}
- i \frac{\der \Gamma}{\der \Psi^\dag}
 \supset i \Im \lkk \frac{\der J^\dagger (x)}{\der \Psi^\dag}
 \int_y  G_{\rm ret} (x, y) J(y) \rkk, 
\end{align}
Here $G_\text{ret} (x,y)$ is the retarded propagator for relativistic modes $\delta \phi$,
which satisfies:\footnote{
	Here we have neglected a small shift of the mass, $m_\phi$, in the presence of the I-ball/oscillon background.
	The effective mass changes according to the spatial gradient of the I-ball/oscillon.
	The non-relativistic condition indicates that the gradient is much larger than $1/m_\phi$.
	Since we are interested in relativistic modes which has energy and momentum comparable to or larger than $m_\phi$,
	we may use this approximation.
}
\begin{align}
	\prn{ \Box_x + m_\phi^2 } G_\text{ret} (x,y) = \delta (x-y).
\end{align}
It is convenient to consider it in the momentum space:\footnote{
	Here and hereafter, 
	we have used the shorthanded notation: $\int_k \equiv \int \dd^4 k / (2\pi)^4$
	and $\int_{\bm k} \equiv \int \dd^3 k / (2\pi)^3$.
}
\begin{align}
 \int_y G_{\rm ret} (x, y) J(y) 
 = 
 \int_K G_{\rm ret} (K) J(K) e^{- i K \cdot x}, 
\end{align}
where 
\begin{align}
 \Im G_{\rm ret} (k) = \frac{\pi}{2 \omega_k} \lkk \delta \lmk k_0 - \omega_k \rmk 
 - \delta \lmk k_0 + \omega_k \rmk \rkk 
 \\ 
 J(K) = (2 \pi) \delta \lmk k_0 - n \omega \rmk 
 \tilde{J}_{n+1} ({\bm k})
 \\
 \tilde{J}_{n+1} ({\bm k}) \equiv 
 \frac{g_{n+1}}{2^{3n/2}} 
 \int_{\bm x} e^{-i {\bm k} \cdot {\bm x}} \rad^n ({\bm x}). 
 \label{tildeJ}
\end{align}
Thus we have the imaginary part of the following term in the equation of motion for $\Psi$:
\begin{align}
	\frac{\der \Gamma}{\der \rad (x)} \supset -
	\Im \com{
	\frac{2n g_{n+1}}{2^{3n/2} 
	}\, \rad^{n-1} (x)\, e^{i n m_\phi x_0} \,
	\int_K G_\text{ret} (K) \, J(K) e^{-iK \cdot x}
	}.
	\label{eq:imaginary}
\end{align}
The imaginary part corresponds to the production of on-shell relativistic modes outside the I-ball/oscillon.
Plugging the I-ball/oscillon solution into $\rad$, one can estimate its decay rate perturbatively.
After some algebras,
we finally obtain the decay rate of the charge in the case of Eq.~\eqref{eq:int_ex}:
\begin{align}
	\dot Q 
	&= 
	- 2 \pi n \int_{\bm{k}} \abs{\tilde{J}_{n+1} (\bm{k})}^2  \delta \lmk n^2 \omega^2 - m_\phi^2 - \abs{\bm{k}}^2 \rmk 
	\\
	&\simeq - \frac{n}{2 \pi} \sqrt{n^2-1} \omega \abs{\tilde{J}_{n+1}}^2, 
	\label{dot Q}
\end{align}
where $\tilde{J}_{n+1}$ is given by \eq{tildeJ} with $\abs{\bm k} = \sqrt{n^2 \omega^2 - m_\phi^2} \simeq \sqrt{n^2 - 1}\omega$. 
In the second line, we use $\psi (\bm{x}) = \psi (r)$ and $\tilde{J}_{n+1} (\bm{k}) = \tilde{J}_{n+1} (\abs{\bm{k}})$.
This is the emission rate of particles 
with a momentum of $\abs{\bm k} = \sqrt{n^2 \omega^2 - m_\phi^2}
\simeq \sqrt{n^2 - 1} \omega$
via the $n+1$-point interaction.

Once we assume the configuration as the Gaussian form of \eq{eq:gaussian}, 
we can analytically calculate $\tilde{J} ({\bm k})$ and obtain 
\begin{align}
	\dot Q = - C(n) \, \tilde g_{n+1}^2 \epsilon^{3(n-2)} 
	\exp \prn{ - \frac{n^2 - 1}{2 n} \frac{1}{\epsilon^2} }\,
	\omega \times Q^n,
	\label{dQdt Gaussian}
\end{align}
where 
the order one factor $C(n)$ are defined as 
\begin{align}
	C(n) \equiv \frac{2^{(n - 2)/2} \sqrt{n^2 - 1}}{ \pi^{3n/2 - 2} n^2}.
\end{align}
Here we have defined a dimensionless coupling, $\tilde g_{n+1} \equiv g_{n+1} \omega^{n-3}$,
and the epsilon parameter, $\epsilon \equiv 1 / \omega R$.
As can be inferred from Eq.~\eqref{eq:imaginary},
this process corresponds to the production of relativistic modes with 
$k_0 \simeq n \omega$ and $k \simeq \sqrt{n^2 - 1} \omega$
owing to the combination of high momentum modes 
inside the I-ball/oscillon
and the U(1) breaking interaction terms.
Note that
the non-relativistic condition indicates that the epsilon parameter should be smaller than unity, $\epsilon \ll 1$, which
ensures that the decay rate is exponentially suppressed.
Also, note that the rate becomes exponentially smaller for a larger $n$, 
which is confirmed in our full numerical simulation as we explain in the next section. (See also Refs.~\cite{Segur:1987mg,Gleiser:2008ty,Fodor:2008du,Fodor:2009kf,Hertzberg:2010yz} for instance.) 
This implies that the cubic term $(n=3)$ makes the life time of the I-ball/oscillon shorter than that of the $\mathbb Z_2$ invariant case.

\subsubsection*{Toy model and its implications}

Here we apply the above theory of decay process to a toy model. 
Let us consider a Q-ball in a U(1) theory with the potential of \eq{V_eff}. 
Once we specify the coupling constants, all properties of Q-ball are determined 
as a function of its charge $Q$. 
Then we impriment its time evolution by imposing \eq{dot Q} by hand. 
This is a toy model that simply describes the time evolution of I-ball/oscillon via the decay process discussed above. 
One can see that the qualitative result of this toy model is consistent with the result of full numerical simulation of I-ball/oscillon that we explain in the next section.

\begin{figure}[t]
\centering 
 \includegraphics[width=.40\textwidth ]{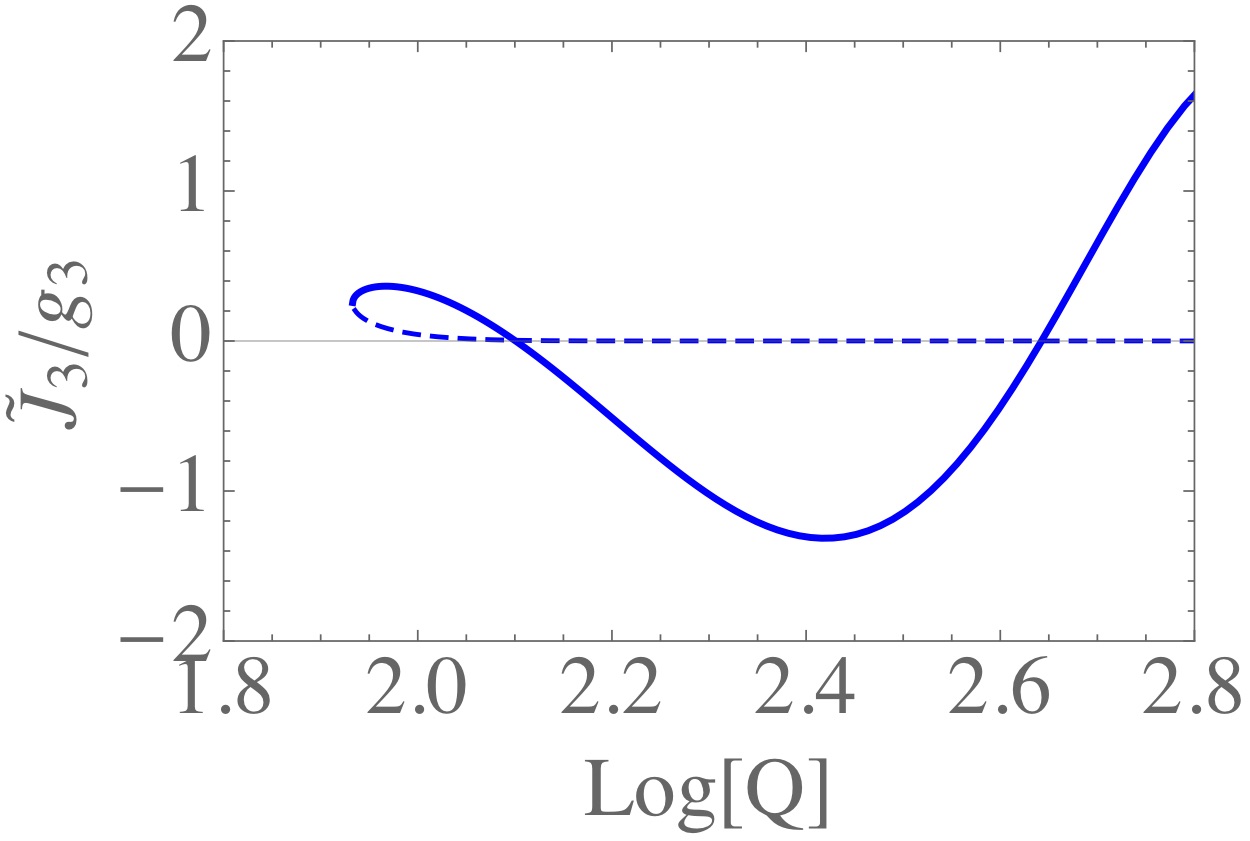} 
 \qquad 
 \includegraphics[width=.40\textwidth ]{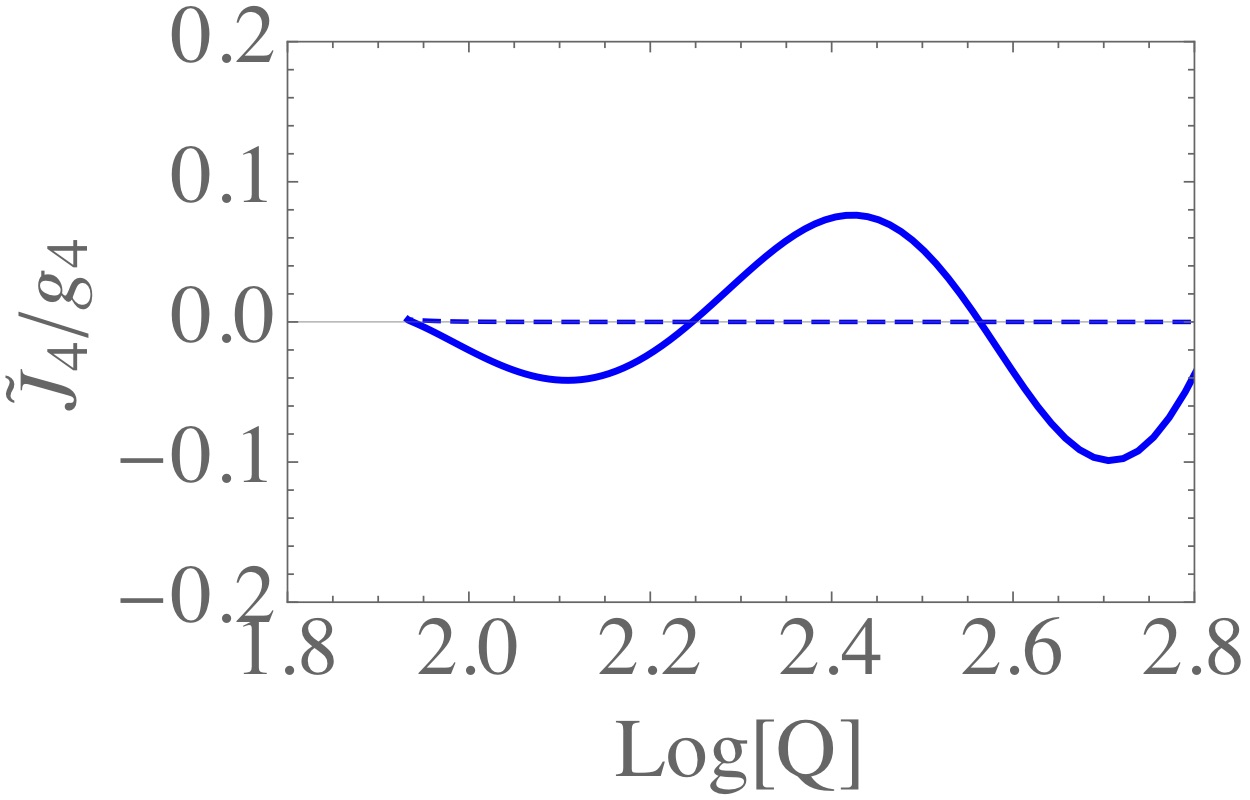}  
 \\
 \includegraphics[width=.40\textwidth ]{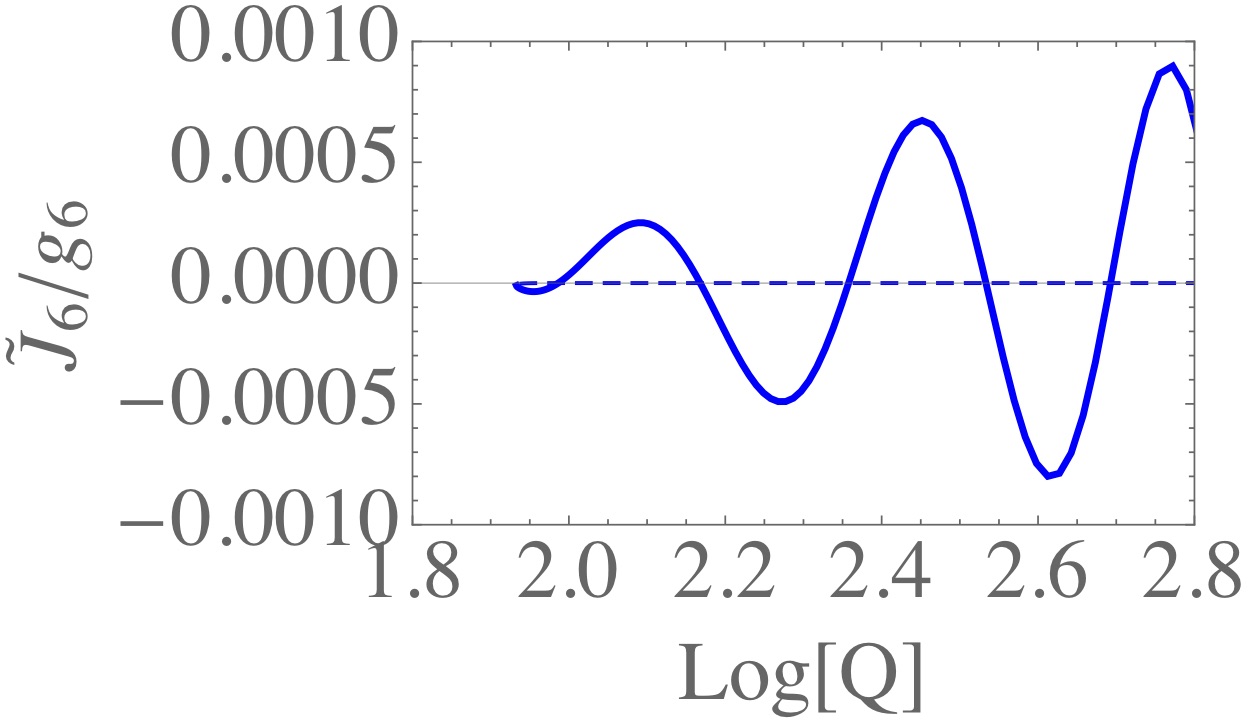} 
 \qquad 
 \includegraphics[width=.40\textwidth ]{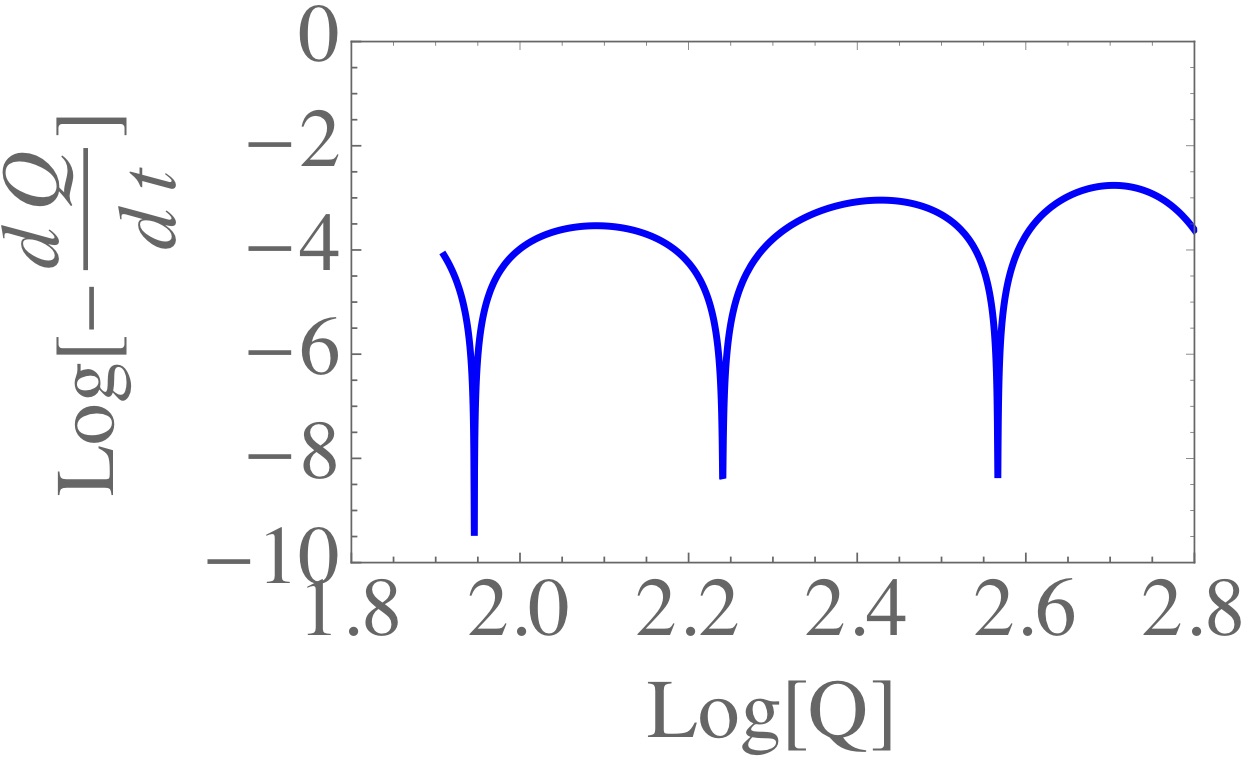}  
\caption{\small
Values of $\tilde{J}_{n}/ g_{n}$ as a function of charge $Q$ for $n = 3$ (upper left panel), 
$n=4$ (upper right panel), and $n = 6$ (lower left panel), 
and the sum of $- dQ / dt$ (lower right panel). 
We assume $g_4 = 1$ and $g_6 = 0.4$. 
}
  \label{fig:J}
\end{figure}

For a given charge $Q$, 
we can calculate its configuration $\rad (r)$ 
and then we can calculate $\tilde{J}_n$ numerically. 
Figure~\ref{fig:J} shows the value of $\tilde{J}_n / g_{n}$ as a function of charge $Q$ 
for $n=3$, $4$, and $6$. 
We are interested in solutions on the stable branch, which is shown as solid blue lines in the figure. 
We confirm that 
typical values of $\tilde{J}_n/ g_{n}$ is exponentially smaller for a larger $n$, 
which is explicitly shown above in the Gaussian profile approximation as Eq.~(\ref{dQdt Gaussian}). 
We can see that $\tilde{J}_n/ g_{n}$ oscillates as a function of charge 
and it vanishes at certain values. 
This implies that the decay process becomes inefficient for a solution with charges around these values. 
For example, 
suppose that the initial charge of I-ball/oscillon is $10^{2.7}$. 
If the three point interaction is absent, that is, if $g_3 = 0$, 
its charge decreases mainly via the four point interaction. 
However, 
when its charge decreases to $10^{2.57}$, 
$\tilde{J}_4^2$ decreases to $0$ and its decay rate is suppressed accordingly. 
This implies that I-ball/oscillon reaches an attractor solution in this toy model. 
In realistic, 
we also have a six-point interaction because of $g_6 = 0.4$, which gives nonzero decay rate 
at the critical value though it is exponentially suppressed. 
Therefore, the attractor solution is not exactly stable 
and the charge can decrease below the point of $\tilde{J}_4 =0$. 
This can be seen in the lower right panel in the figure, where 
we plot the value of $-dQ/dt$ with the contributions of four and six point interactions. 
Although the typical value of decay rate is of order $10^{-2}$, 
it is as small as $10^{-8}$ but is still nonzero at the attracter points. 
Eventually, the solution goes away from the attractor point 
and then 
$\tilde{J}_4$ increases and the decay rate increases as shown in the figure. 
We can see that there are also points at which $\tilde{J}_4 =0$ around $Q = 10^{2.24}$ 
and $Q = 10^{1.95}$. 
This implies that the time-evolution of charge has many bumps and attractor region 
due to these properties. 
Note that there is a point at which $\tilde{J}_4 =0$ just above the critical point $Q_{\rm cr}$. 
This implies that 
the I-ball/oscillon reaches an attractor and then suddenly decays due to the classical instability. 
This feature is checked numerically as we explain in the next section.

\begin{figure}[t]
\centering 
 \includegraphics[width=.60\textwidth ]{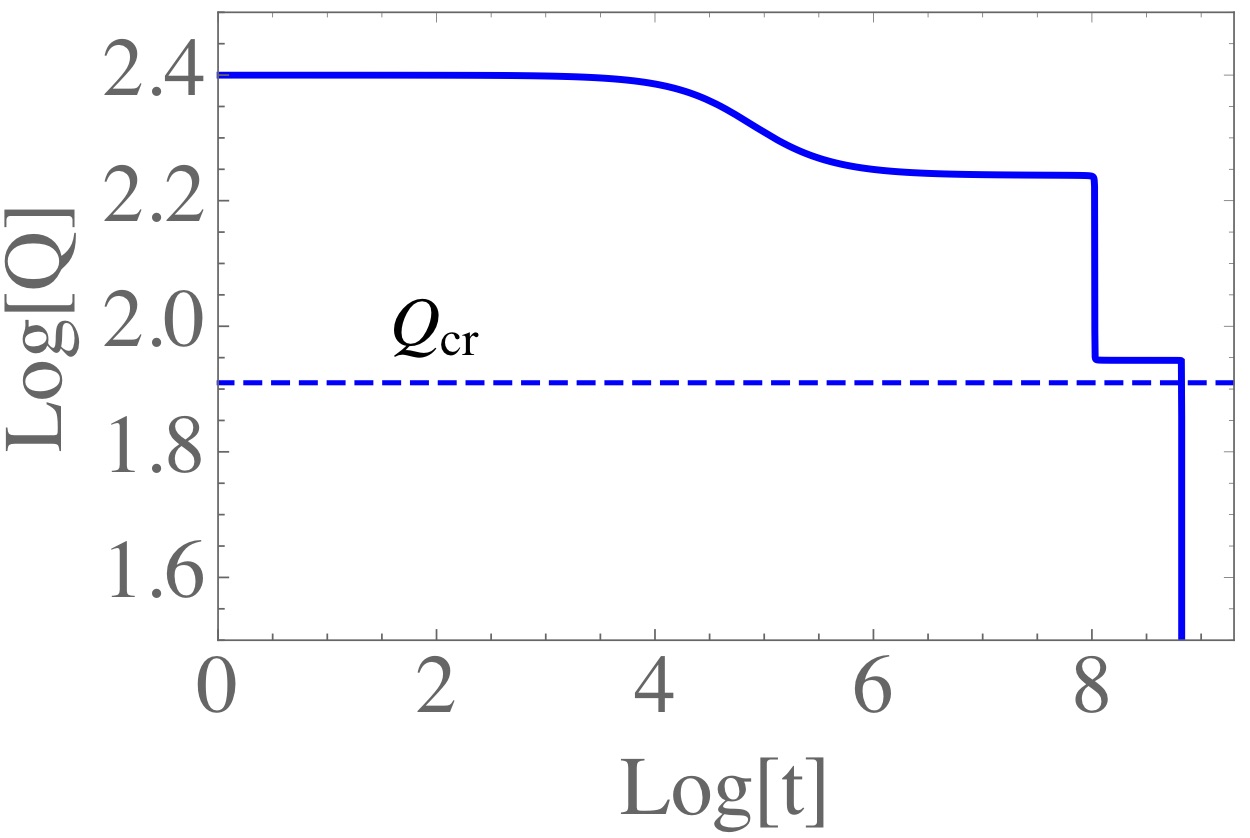} 
\caption{\small
Time evolution of charge $Q$ in the toy model. 
We assume $g_4 = 1$ and $g_6 = 0.4$. 
The blue dashed line represents the critical value $Q_{\rm cr} \simeq 10^{1.91}$ below which 
there is no solution of I-ball/oscillon. 
}
  \label{fig:evolution-toy}
\end{figure}

At each time step, 
we can calculate its configuration 
for a given charge $Q$. 
Then we can calculate $\tilde{J}_n$ numerically 
and thus we obtain its decay rate \eq{dot Q}, 
which determines charge in the next time step. 
In this way, we can obtain the time evolution of these quantities. 
Figure~\ref{fig:evolution-toy} shows the time evolution of charge $Q$ in this toy model, 
where we assume $g_3 = 0$ and the initial value of its charge is taken to be $10^{2.4}$. 
Since the initial charge is smaller than $10^{2.57}$, 
it decreases fast due to the four-point interaction. 
Then it reaches to an attractor point, at which the charge is about $10^{2.24}$. 
After that, the decay rate is suppressed as low as $10^{-8}$ 
but is still nonzero (see Fig.~\ref{fig:J}). 
Then it escapes from the attractor point and its charge decreases again 
due to the four-point interaction. 
Then it reaches 
the second attractor point at $Q \simeq 10^{1.95}$, which is just above the critical value $Q_{\rm cr} \simeq 10^{1.91}$. 
This means that 
there is some time before the I-ball/oscillon disappears due to the classical instability. 
We compare these results with the full numerical simulation of 
the evolusion of I-ball/oscillon in the next section.

\section{Numerical Simulation}
\label{sec:num_sim}

We perform numerical simulations to solve the equation of motion of a real scalar field 
in 3+1 dimensions, assuming spherical symmetry to reduce the spacial dependence in 3 dimensions to 
a radial dependence. 
In this section, we show our results of numerical simulations 
that justify our theory discussed in the previous section.

We can rescale parameters and variables 
such as 
\beq
 t &\to& t / m_\phi
 \\
 \phi &\to&  \phi / \sqrt{g_4} 
 \\
 r &\to& r / m_\phi
\eeq
to eliminate $m_\phi$ and $g_4$ in the real scalar field theory. 
In other words, we can use units of $m_\phi = 1$ and $g_4 = 1$ in \eq{V_phi} without loss of generality. 
The equation of motion for a scalar field with a polinomial potential is then given by 
\beq
&& \frac{\dd^2}{\dd t^2} \phi - 
 \frac{\dd^2}{\dd r^2} \phi - 
 \frac{2}{r} \frac{\dd}{\dd r} \phi +
 \frac{\del }{\del \phi} V_{\rm num} (\phi) = 0, 
 \\
 &&V_{\rm num} (\phi) = 
 \frac{1}{2} \phi^2 + 
 \frac{1}{3} g_3 \phi^3 - 
 \frac{1}{4} \phi^4 + 
 \frac{1}{6} g_6 \phi^6, 
 \label{V_num}
\eeq
where we assume spherical symmetry 
and consider the dynamics of radial direction. 
Here we redefine $g_3$ and $g_6$ to absorb $m_\phi$ and $g_4$ after rescaling the variables and the field value.

We develop a numerical lattice code to solve the classical equation of motion 
and investigate the time-evolution of I-ball/oscillon. 
We use the 6th order leapfrog method for time-evolution. 
We take the lattice grid number as $N = 10^3$ 
and 
the lattice volume as $L = 50$ (in the unit of $m_\phi = 1$), 
which is much larger than a typical size of I-balls/oscillons. 
The grid size is thus given by $\Delta x = L / N = 0.05$. 
We take each time step as $\Delta t = 0.01$ 
and numerical simulation is performed from $t = 0$ to $t = T_{\rm max} = 10^5$. 
We check that 
our results of numerical simulations are rarely dependent 
on these numerical setups. 
For example, 
the heights and peak locations in Fig.~\ref{fig:spectrum1} 
do not change 
when we change $N$, $L$, and/or $\Delta t$ by amounts of $50 \%$. 

The boundary condition at $r = 0$ is taken to be $\del \phi / \del r (r = 0) = 0$ 
while 
the one at the other boundary (\textit{i.e.}, at $r= L$) is taken to be 
the absorbing boundary condition (see App. A in Ref.~\cite{Salmi:2012ta}).%
\footnote{
There is a typo in Eq.~(A9) in Ref.~\cite{Salmi:2012ta}. 
The sign of the third tern $-1/2 m^2 \cphi$ should be positive for a correct absorbing boundary condition. 
}
The absorbing boundary condition allows us to investigate 
the time-evolution of I-ball/oscillon 
less affected by the boundary effect at $r = L$ 
because emitted particles from the I-ball/oscillon are absorved and are rarely reflected by that boundary. 
We are interested in the time-evolution of I-balls/oscillons after they form. 
Thus, initial conditions are taken to be 
a spacially-localized gaussian configuration: 
\beq 
 \phi (t=0, r) = \phi_{\rm ini} e^{- r^2 / R_{\rm ini}^2}, 
\eeq
and $\dot{\phi} = 0$, where $\phi_{\rm ini}$ and $R_{\rm ini}$ are some constants specified below.

Note that some important quantities discussed in the previous section, such as 
$Q$ and $R_Q$, 
cannot be directly deduced from our numerical simulation 
because they are defined by $\Psi$ (not $\phi$) [see \eq{eq:charge}]. 
In order to compare our numerical results with our theory discussed in the previous seciton, 
we approximate them by taking time-avarage over some period of oscillation-time scale $m_\phi^{-1}$. 
For example, 
$Q$ is approximated by $\bar{Q}$: 
\beq
 \bar{Q} = 
\frac{1}{T_{\rm avl}} \int_{t-T_{\rm avl}}^{t} \dd t \int \dd^3 x  \dot{\phi}^2 
\label{barQ}
 \\
 = \frac{1}{T_{\rm avl}} \int_{t-T_{\rm avl}}^{t} \dd t \int 4 \pi  r^2 \dot{\phi}^2 \dd r, 
\eeq
and $R_Q$ is approximated by $\bar{R}_Q$: 
\beq
 \bar{\psi}(r = \bar{R}_Q) = \bar{\psi}(r = 0) / e, 
 \label{barR}
 \\
 \bar{\psi} (t, r) \equiv 2 \lkk \frac{1}{T_{\rm avl}} \int_{t-T_{\rm avl}}^{t} \dd t \phi^2 (t, r) \rkk^{1/2}, 
 \label{barphi}
\eeq
where $T_{\rm avl}$ is a duration time for time-avarage. 
We replace the integral by a summation 
in our numerical simulations. 
We take $T_{\rm avl} = 10$ in our numerical simulation. 
On the other hand, 
the energy is defined by $\phi$ such as \eq{eq:energy} without taking time-avarage.

\subsection{Case with $\mathbb{Z}_2$ symmetry}

\begin{figure}[t]
\centering 
 \includegraphics[width=.40\textwidth ]{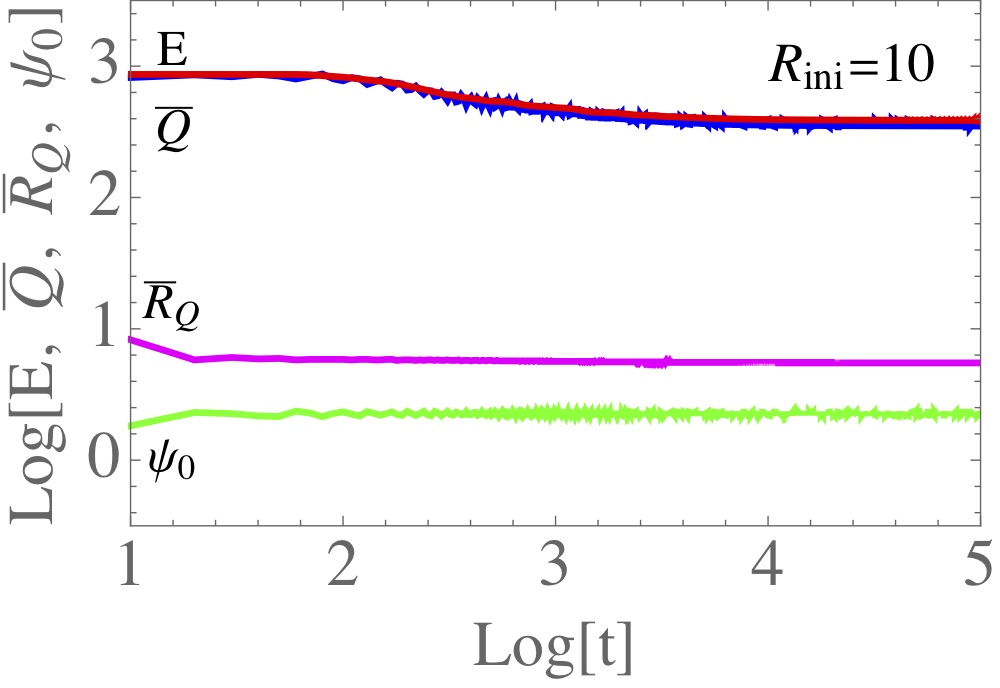} 
 \qquad 
 \includegraphics[width=.40\textwidth ]{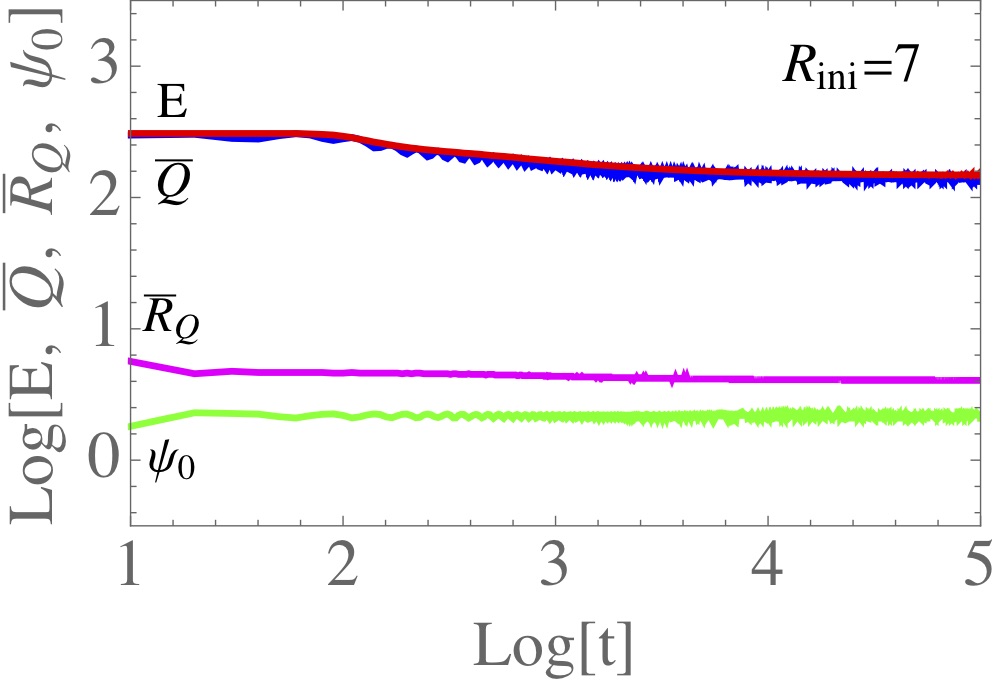} 
 \\
 \vspace{0.5cm}
 \includegraphics[width=.40\textwidth ]{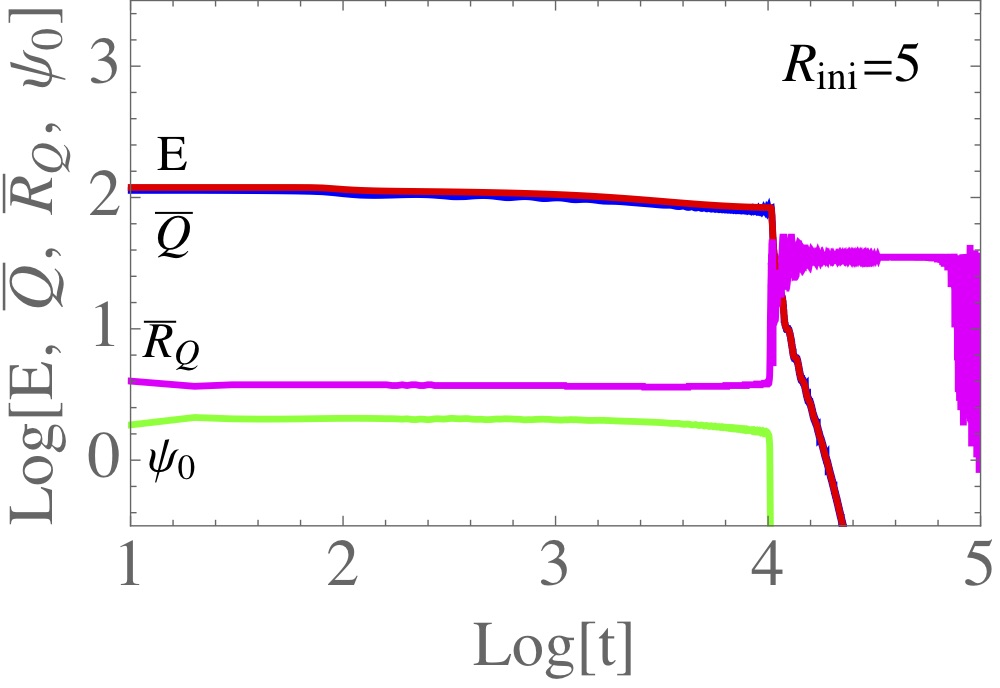} 
 \qquad 
 \includegraphics[width=.40\textwidth ]{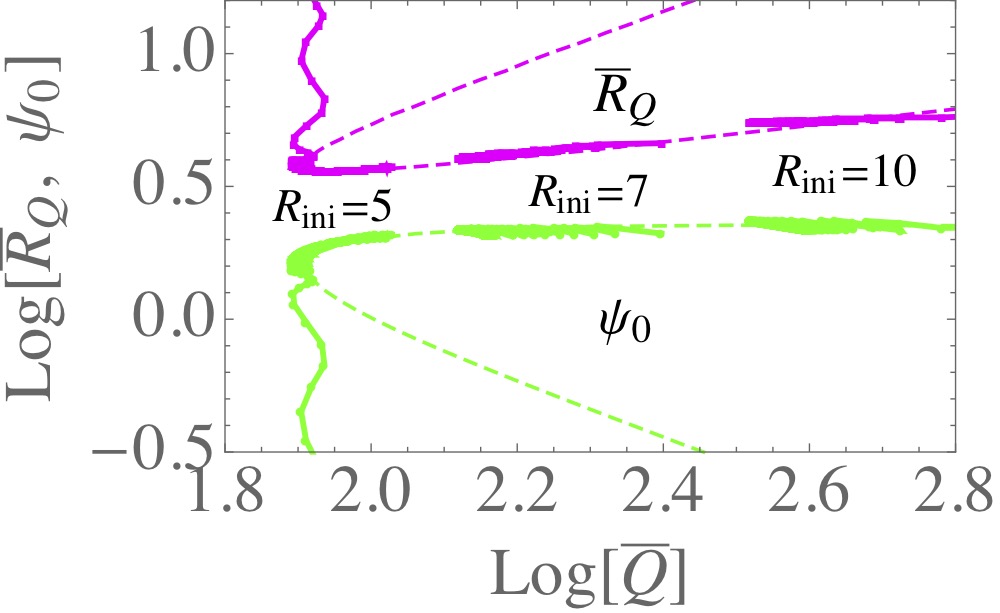}  
\caption{\small
Time-evolution of $E$ (red line), $\bar{Q}$ (blue line), $\psi_0$ (green line), and $\bar{R}_Q$ (magenta line) as a function of time for $R_{\rm ini} = 10$ (upper left panel), 
$R_{\rm ini} = 7$ (upper right panel), $R_{\rm ini} = 5$ (lower left panel), 
and that of $\psi_0$ (green line) and $\bar{R}_Q$ (magenta line) as a function of $\bar{Q}$ for $R_{\rm ini} = 5, 7, 10$ (lower right panel). 
In the lower right panel, 
we also plot theoretical lines 
of $\psi_0$ (green dashed line) and $R_Q$ (magenta dashed line). 
We take $g_6 = 0.4$ and $\phi_{\rm ini} = 1$. 
We also plot theoretical lines of $R_Q$ and $\psi_0$ (\textit{i.e.}, $\rad_0$) as a function of $Q$ (dashed lines), 
which are computed as explained
in the previous section
[See also \eq{V_eff} and \eq{eq:charge}].
}
  \label{fig:evolution1}
\end{figure}

Here we show our results of numerical simulation for the case with $\mathbb{Z}_2$ symmetry (\textit{i.e.}, 
for the case of $g_3 = 0$). 
We take $g_6  = 0.4$ as an example. 
Figures~\ref{fig:evolution1} show the time-evolution of 
$\psi_0$, $\bar{R}_Q$, $E$, and $\bar{Q}$ as a function of time 
for $R_{\rm ini} = 10$ (upper left panel), 
$R_{\rm ini} = 7$ (upper right panel), 
and $R_{\rm ini} = 5$ (lower left panel), 
and the contours of 
$\psi_0$ and $\bar{R}_Q$ 
as a function of $\bar{Q}$ (lower right panel). 
Here, $\psi_0$ is the time-averaged field value at $r= 0$ (\textit{i.e.}, $\psi_0 \equiv \bar{\psi} (r = 0)$) 
and $\bar{R}_Q$ is a typical size of I-ball/oscillon defined by \eq{barR}. 
We take $\phi_{\rm ini} = 1$. 
We divide the time interval $[0, T_{\rm max}]$ by $10^4$ segments 
and take average over each segment to calculate $\psi_0$, $\bar{R}_Q$, and $\bar{Q}$ 
(\textit{i.e.}, we take $T_{\rm avl} = T_{\rm max} / 10^4 = 10$.) 
This is the reason that the plot starts at $t = T_{\rm max} /10^4 = 10$ in the figures. 

From the upper left panel, we can see that 
the averaged charge $\bar{Q}$ reaches an attractor point 
and the I-ball/oscillon is too long-lived to calculate its time evolution numerically until it reaches the critical point. 
Although the I-ball/oscillon in the upper right panel starts from an averaged charge below 
that attractor point, it reaches another attractor point and again it is long-lived. 
The I-ball/oscillon in the lower left panel starts from an averaged charge 
below the latter attractor point and can reaches the critical value, at which 
the I-ball/oscillon suddenly decay due to the classical instability.

We take some initial conditions by changing $R_{\rm ini}$ 
to find time-evolutions of these parameters in 
$\psi_0-\bar{Q}$ and $\bar{R}_Q - \bar{Q}$ planes (the lower right panel in Fig.~\ref{fig:evolution1}). 
We also plot theoretical lines of $R_Q$ and $\psi_0$ (\textit{i.e.}, $\rad_0$) as a function of $Q$, 
which are calculated by the procedure discussed in the previous section 
with \eq{V_eff} [see also \eq{eq:charge}]. 
One can see that our predictions are consistent with the results of numerical simulations. 
Even if we take different value of $R_{\rm ini}$, 
values of $\bar{R}_Q$ and $\psi_0$ soon reach the theoretical lines (before $t = T_\text{max}/10^3$) 
and then evolve along with these lines. 
In particular, 
once $\bar{Q}$ decreases down to a critical point, 
which is defined by $\del \omega / \del Q = 0$ [see \eq{eq:collapse}], 
the I-ball/oscillon suddenly decays.

\begin{figure}[t]
\centering 
 \includegraphics[width=.60\textwidth ]{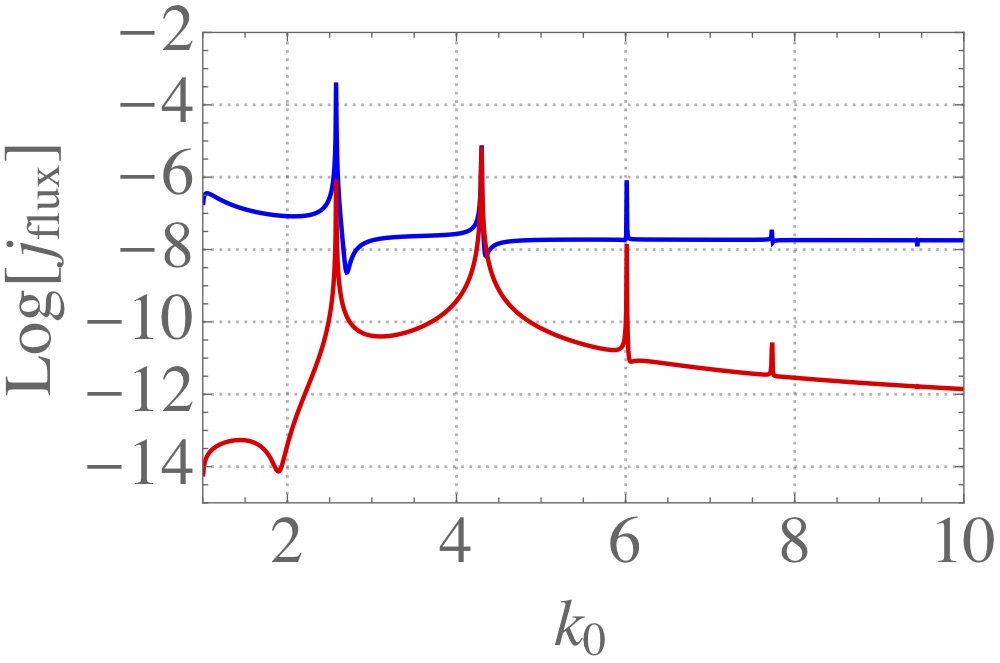} 
\caption{\small
Spectra of energy flux outside I-ball/oscillon (red line) 
and inside I-ball/oscillon (blue line) 
at the time of $t_1 = 10^5$ 
for the case of $g_6 = 0.4$ and $R_{\rm ini} = 10$. 
}
  \label{fig:spectrum1}
\end{figure}

We also plot spectra of particles in Fig.~\ref{fig:spectrum1}, 
where the red line represents the spectrum of energy flux from the I-ball/oscillon 
and the blue one represents that inside the I-ball/oscillon. 
The spectrum is deduced from the numerical simulations 
by extracting field configurations in a certain time interval $[ t_1 -  N_1 \Delta t, t_1]$ 
at a certain radius $r_1$ ($= \Delta x$ or $L$) 
and taking Fourier transformation such as 
\beq
 j_{\rm flux} (k_0) = 4 \pi r_1^2 v \frac{1}{2} k_0^2 \tilde{\phi}^2 (k_0), 
\eeq
where $k_0 = (k^2 + 1^2)^{1/2}$ is energy, $k$ is momentum, 
$v = k/k_0$ is velocity, and $\tilde{\phi} (k_0)$ is Fourier-tramsformed configuration 
calculated by 
\beq
 \tilde{\phi} (k_0) = \frac{1}{N_1} \sum_{i=1}^{N_1} \phi (t = t_1 - N_1 \Delta t + (i-1) \Delta t, r=r_1) e^{2 \pi i (i-1)(j-1)/N_1}, 
\eeq
with $k_0 = 2 \pi (j -1) / (N_1 \Delta t)$ ($j = 1, 2, 3, \dots, N_1$). 
In the figure, we take $t_1 = T_{\rm max} = 10^5$ and $N_1 \Delta t = 10^3$. 
Note that 
the unit energy that each "particle" inside I-ball/oscillon has is not $m_\phi$ but $\omega$ ($= m_\phi - \mu$) [see \eq{eq:omega}], 
which is of order but less than $m_\phi$. 
One can see that I-balls/oscillons dominantly emit relativistic particles with the energy of $(2n+1) \omega$ ($n=1,2,3, \dots$) 
in the case with $\mathbb{Z}_2$ symmetry. 
This is just the process we discuss in Sec.~\ref{sec:decay}. 
Note that we plot the spectrum of particle at the time of $t = T_{\rm max}$, 
at which the I-ball/oscillon stays at the attractor point. 
This implies that the four-point interaction process is suppressed 
and the dominant contribution is due to the six-point interaction. 
This is explicitly shown in the figure, where the dominant contribution of flux is 
the particle with the energy of $5 \omega$.

\subsection{Case without $\mathbb{Z}_2$ symmetry}

\begin{figure}[t]
\centering 
 \includegraphics[width=.30\textwidth ]{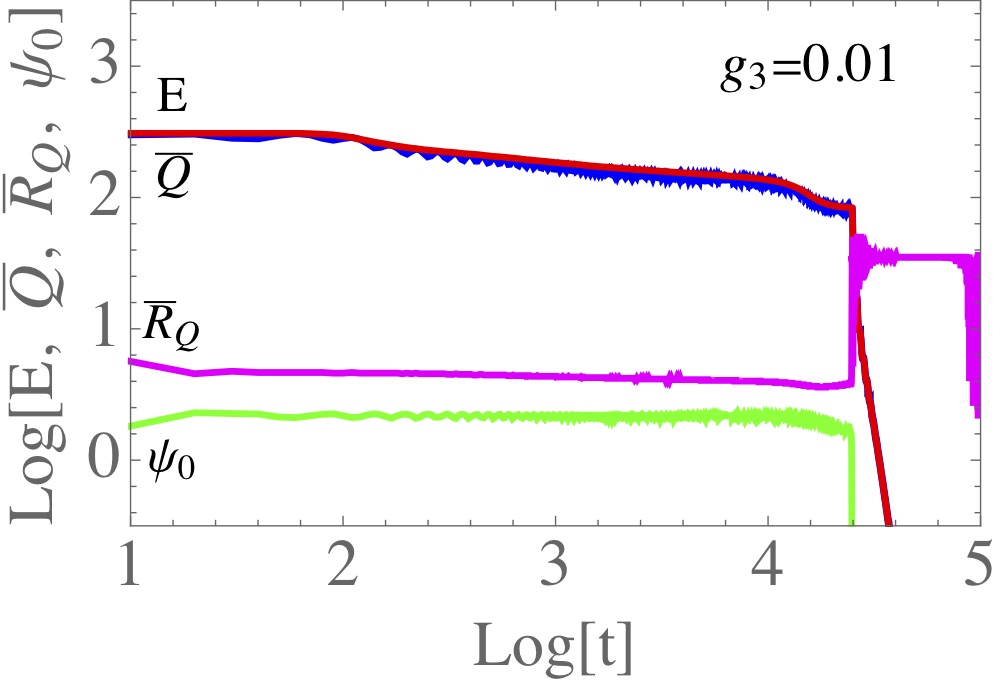} 
 \qquad 
 \includegraphics[width=.30\textwidth ]{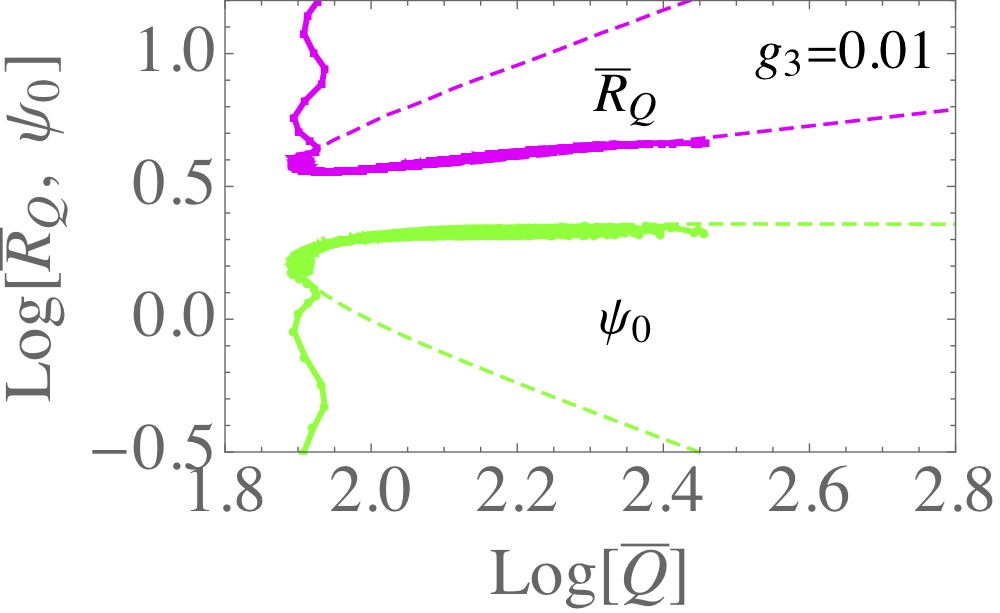}  
 \\
 \vspace{0.2cm}
  \includegraphics[width=.30\textwidth  ]{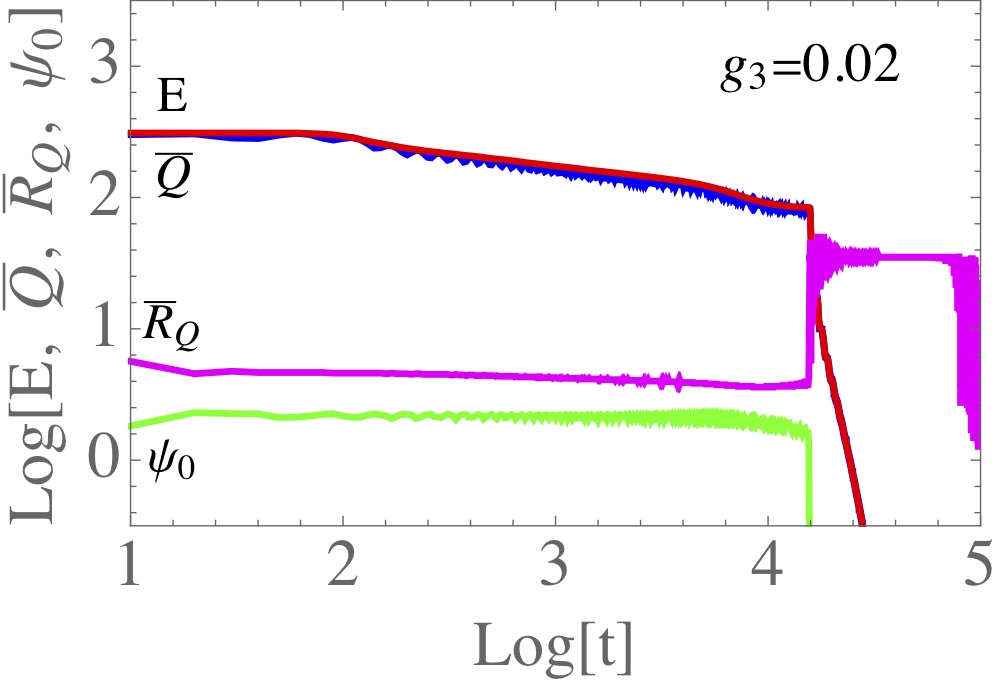} 
 \qquad 
 \includegraphics[width=.30\textwidth ]{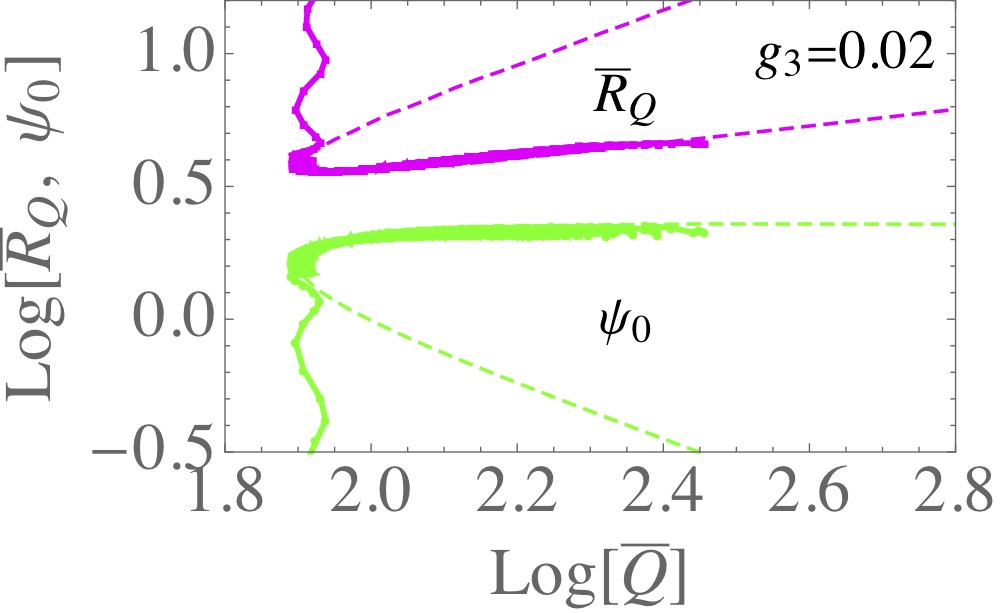}  
\\ 
 \vspace{0.2cm}
\includegraphics[width=.30\textwidth ]{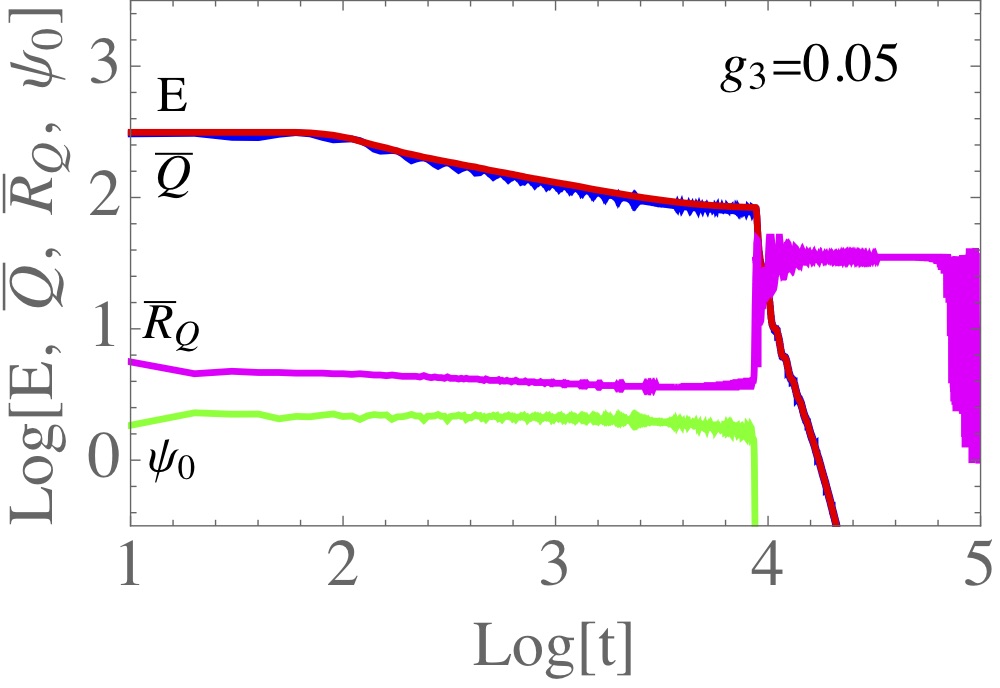} 
 \qquad 
 \includegraphics[width=.30\textwidth ]{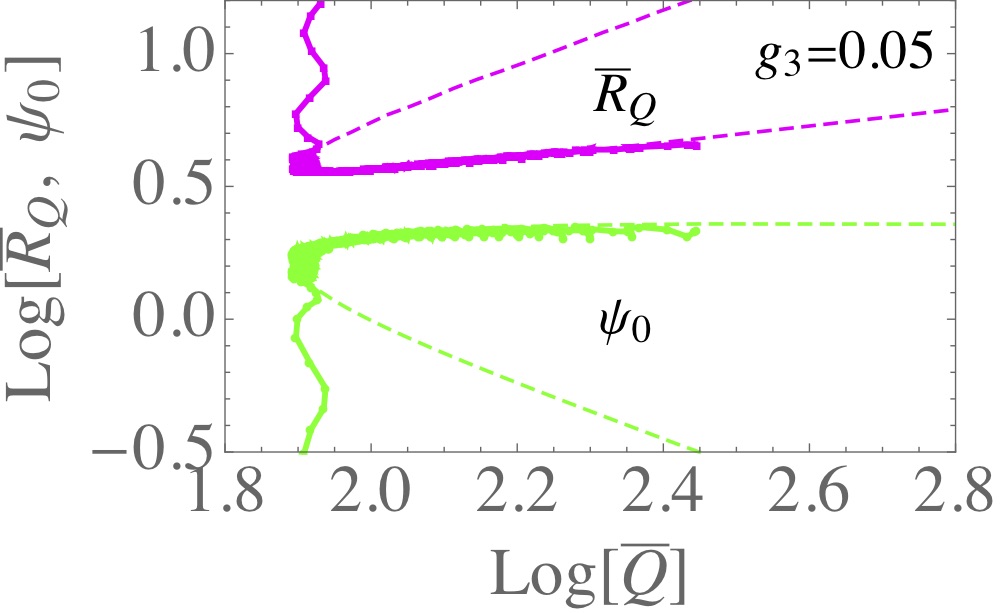}  
\\
 \vspace{0.2cm}
 \includegraphics[width=.30\textwidth ]{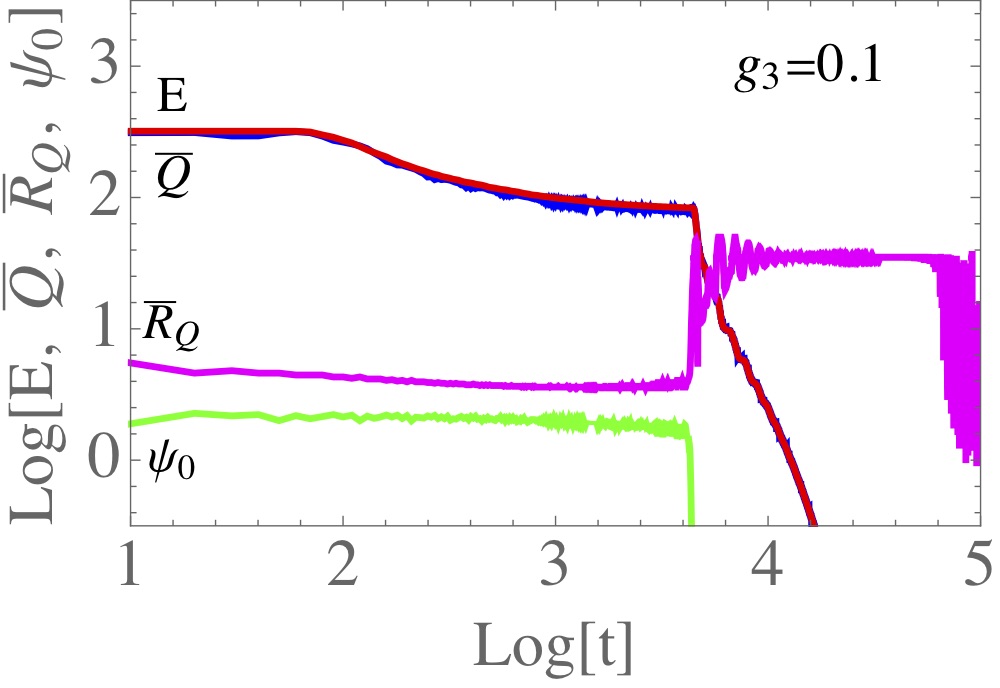} 
 \qquad 
 \includegraphics[width=.30\textwidth ]{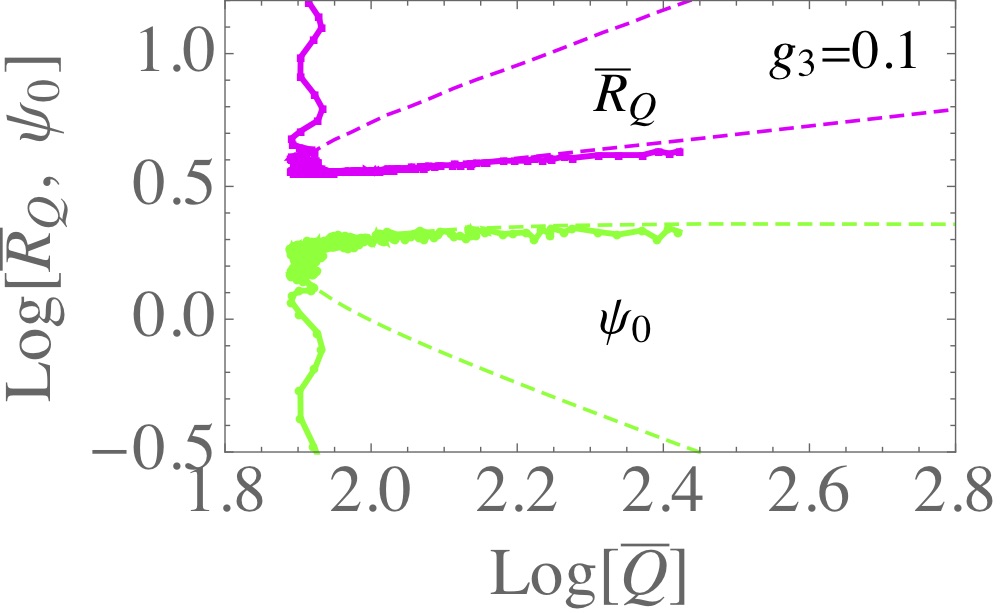}  
\caption{\small
Same as Fig.~\ref{fig:evolution1} but with $g_3= 0.01, 0.02, 0.05, 0.1$. 
}
  \label{fig:evolution2}
\end{figure}

Next we show our results of numerical simulation for the case without $\mathbb{Z}_2$ symmetry. 
Figs.~\ref{fig:evolution2} show the time-evolution of 
$\psi_0$, $\bar{R}_Q$, $E$ and $\bar{Q}$ as a function of time (left panel) 
and $\psi_0$ and $\bar{R}_Q$ 
as a function of $\bar{Q}$ (right panel). 
We take $\phi_{\rm ini} = 1$, $R_{\rm ini} = 7$, 
$g_6 = 0.4$, and $g_3 = 0.01, 0.02, 0.05, 0.1$ as examples. 
One can see that 
the $\mathbb{Z}_2$ breaking term 
affects the evolution of I-ball/oscillon 
and makes I-ball/oscillon decay fast. 
Note that the averaged charge of I-ball/oscillon has approximated attractor points 
for the case of $g_3= 0.01$ (and $0.02$), while there are no such points 
for $g_3 = 0.05$ and $0.1$. 
This implies that the dominant decay process is 
the four-point interaction for the former case while 
it is the three-point interaction for the latter case.

\begin{figure}[t]
\centering 
 \includegraphics[width=.40\textwidth ]{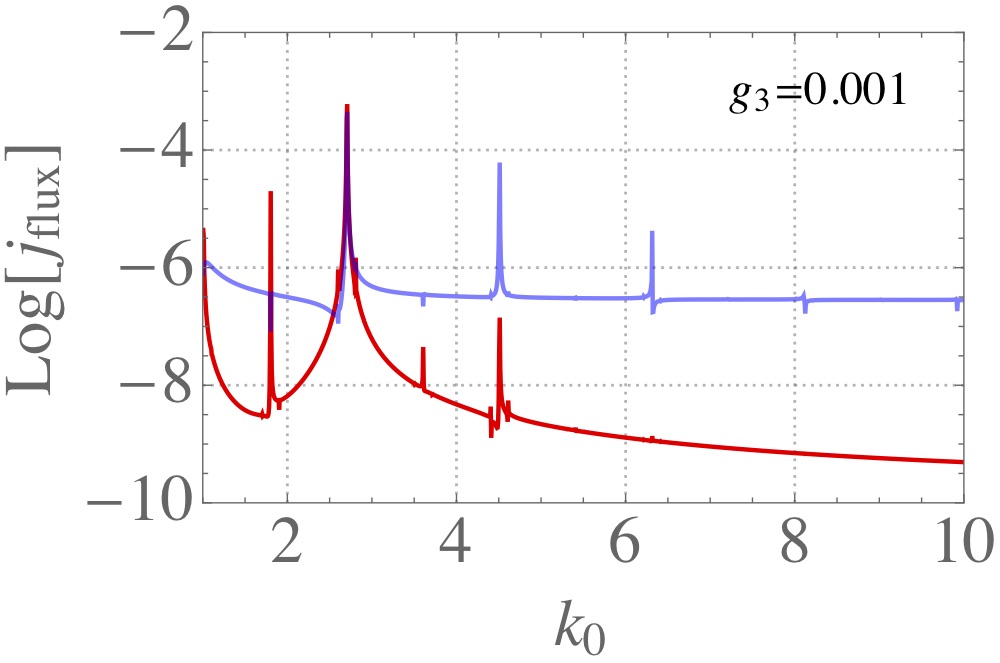} 
 \qquad
 \includegraphics[width=.40\textwidth ]{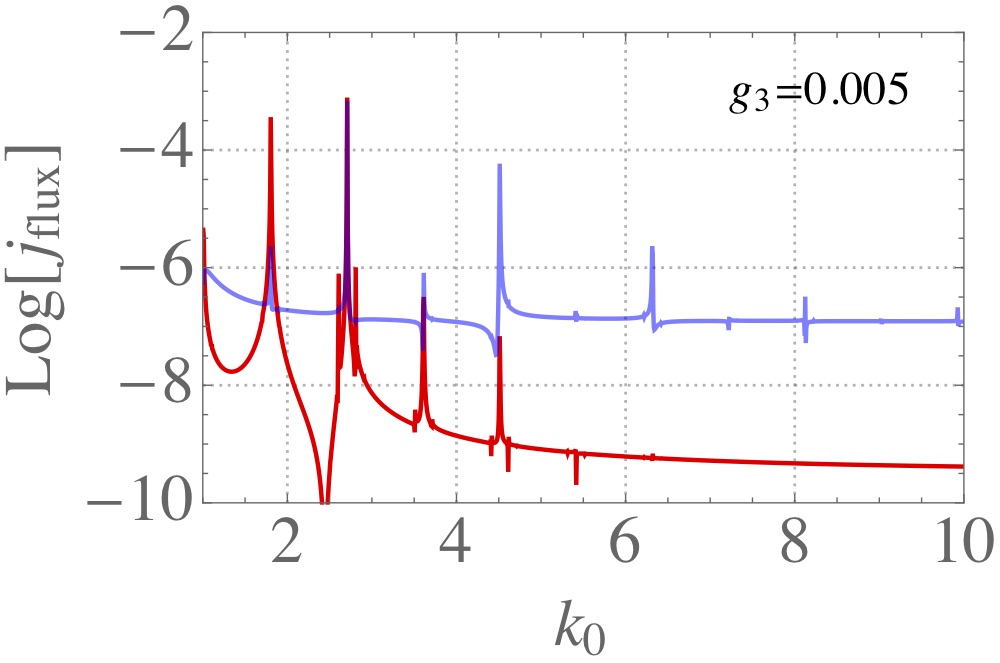} 
\\
 \vspace{0.2cm}
 \includegraphics[width=.40\textwidth ]{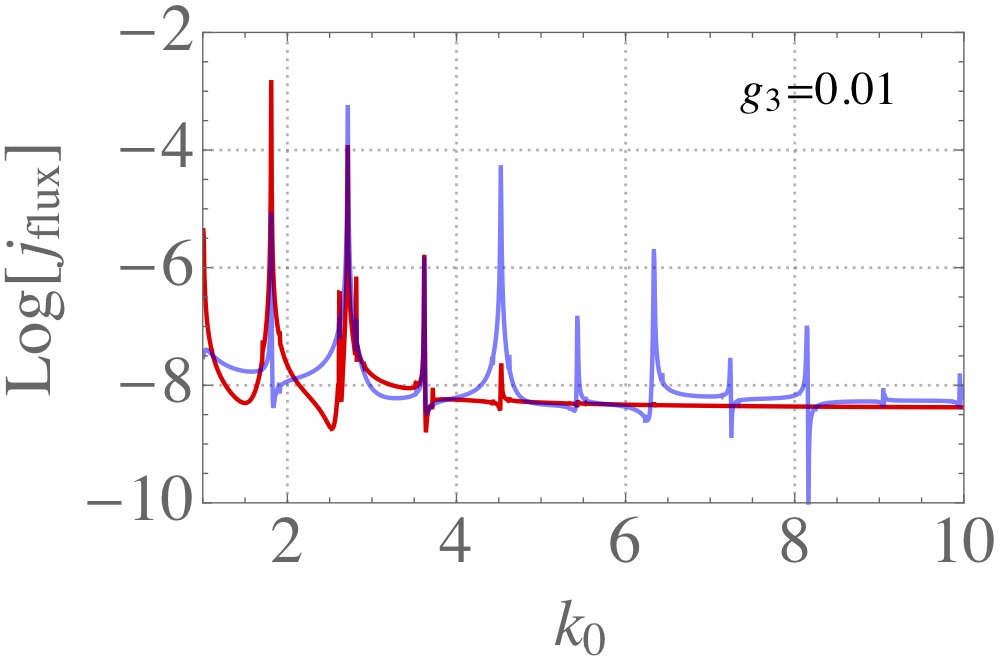} 
 \qquad
 \includegraphics[width=.40\textwidth ]{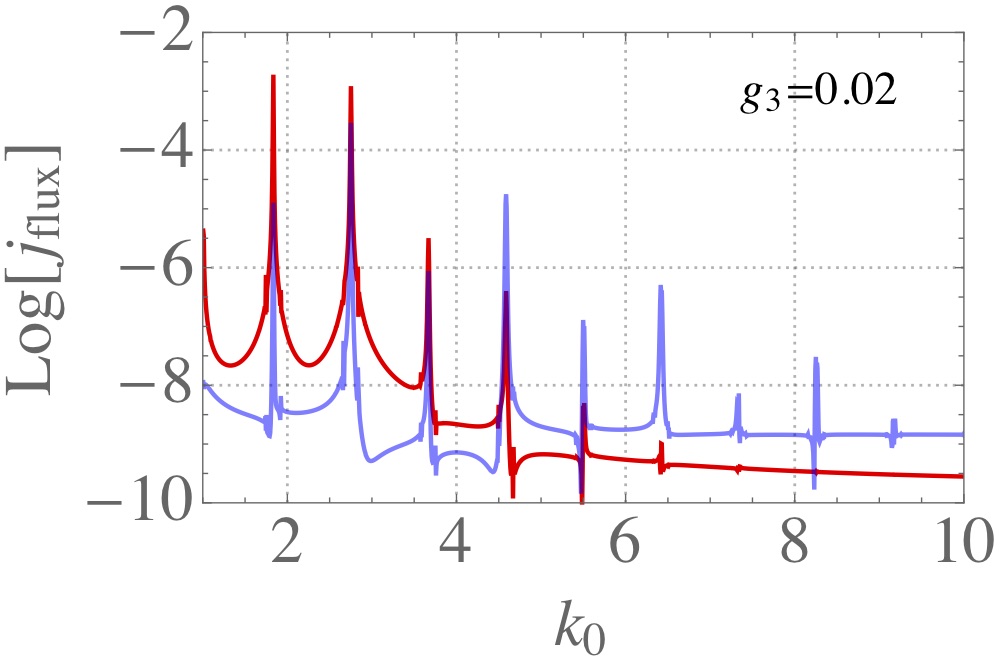} 
\\
 \vspace{0.2cm}
 \includegraphics[width=.40\textwidth ]{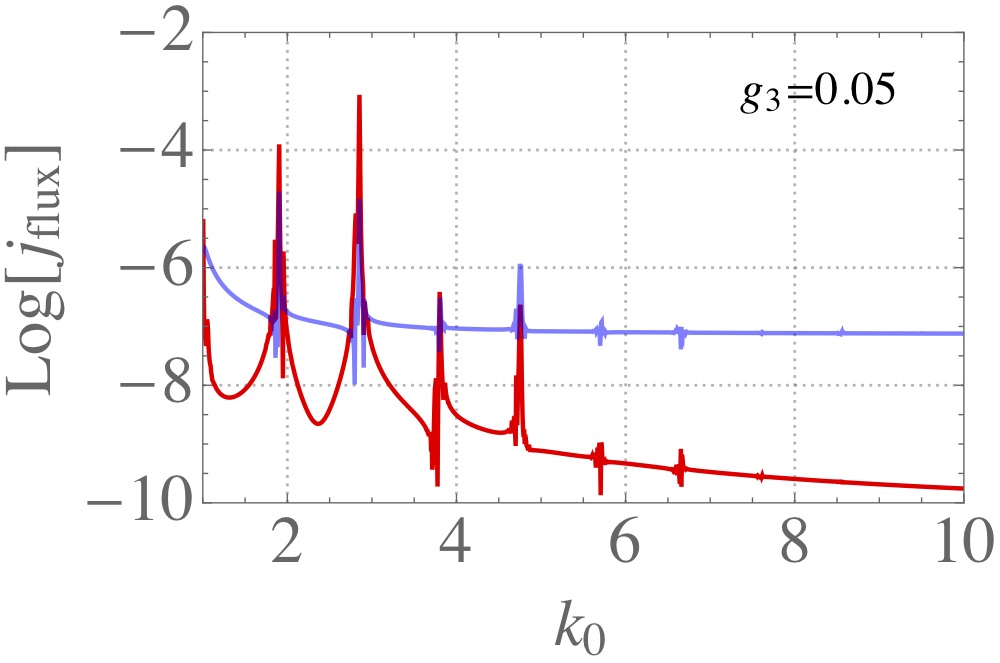} 
 \qquad
 \includegraphics[width=.40\textwidth ]{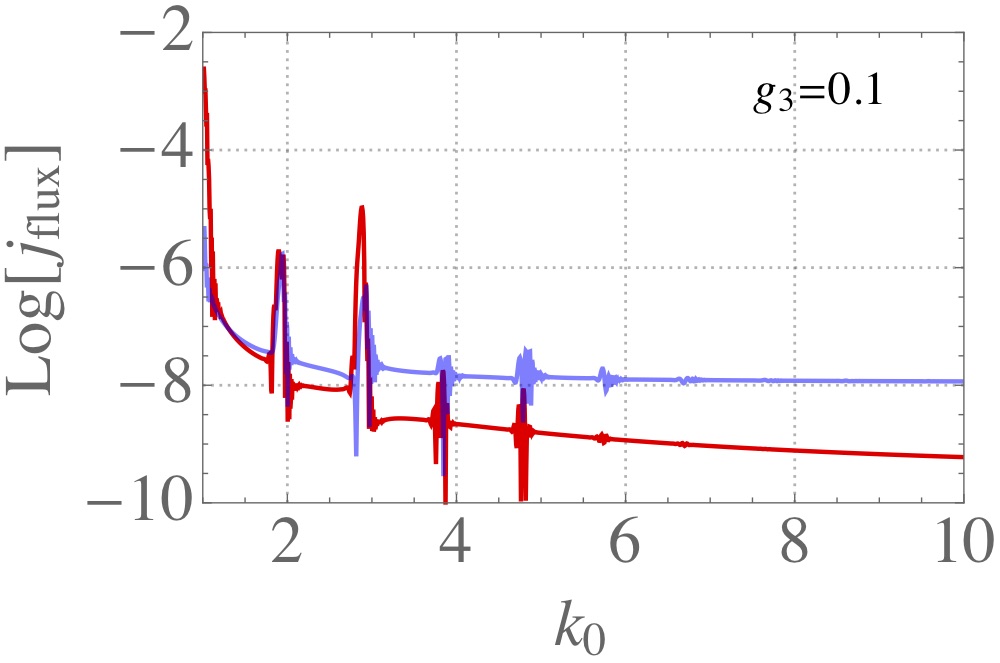}  
\caption{\small
Same as Fig.~\ref{fig:spectrum1} but with $g_3= 0.001, 0.005, 0.01, 0.02, 0.05, 0.1$. 
}
  \label{fig:spectrum2}
\end{figure}

Figure~\ref{fig:spectrum2} 
shows plots of energy flux from I-balls/oscillons at the time of $t_1 = 5 \times 10^3$. 
The $\mathbb{Z}_2$ breaking term leads to decay modes with even units of energy, such as $2 \omega$. 
For $g_3 \lesssim 0.01$, 
the energy flux of even modes is less than that of $3 \omega$, 
while 
for larger $g_3$, 
the energy flux of even modes is comparable or larger than that of $3 \omega$. 
One can also see that 
the height of $2 \omega$ mode 
is roughly proportional to $g_3^2$ for $g_3 \lesssim 0.01$, 
which justifies \eq{dot Q} [see also \eq{dQdt Gaussian}]. 
At the time of $t_1 = 5 \times 10^3$, 
the averaged charge of the I-ball/oscillon is just above the critical value 
for the case of $g_3 = 0.1$. 
This may be the reason why the flux is dominated by the non-relativistic mode 
for $g_3 = 0.1$.

\subsection{Case of double well potential \label{double well}}

Finally, we consider a real scalar field theory with a double well potential, 
which is widely studied in the literature. 
It has a $\mathbb{Z}_2$ symmetry that is spontaneously broken at the minimum of the potential. 
Thus, this case falls into ``without $\mathbb{Z}_2$ symmetry'' in our classification.
The potential is given by 
\beq
 V(\phi) = \frac{1}{4} \lkk \lmk \phi - \frac{\sqrt{2}}{2} \rmk^2 - \frac{1}{2} \rkk^2, 
\eeq
which is a particular case of the one without $\mathbb{Z}_2$ symmetry 
such as $g_3 = 3\sqrt{2}/2$, $g_4 = 1$, and $g_ 6 =0$ 
in \eq{V_phi}. 
Note that we take $\phi = 0$ at the minimum of the potential. 
In this case, the I-ball/oscillon is relatively short lived, 
so that we take $T_{\rm max} = 10^4$ (with $T_{\rm avl}$ fixed) in our numerical simulation. 
We integrate \eq{barQ} from $r = 0$ to $10$ so that 
emitted particles are not included. 

\begin{figure}[t]
\centering 
 \includegraphics[width=.40\textwidth ]{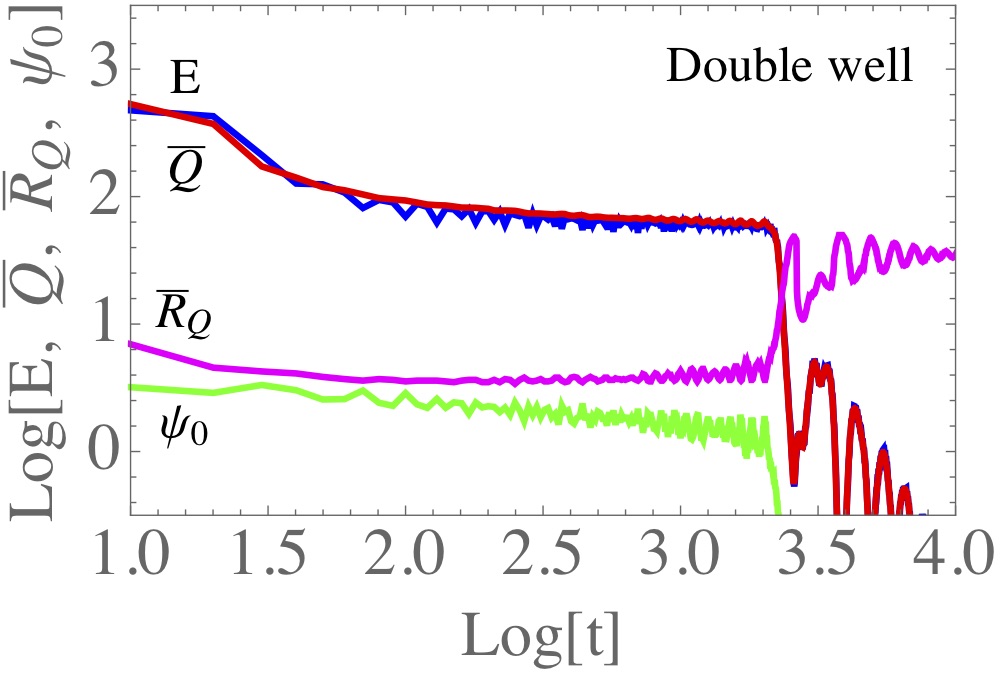} 
 \qquad 
 \includegraphics[width=.40\textwidth ]{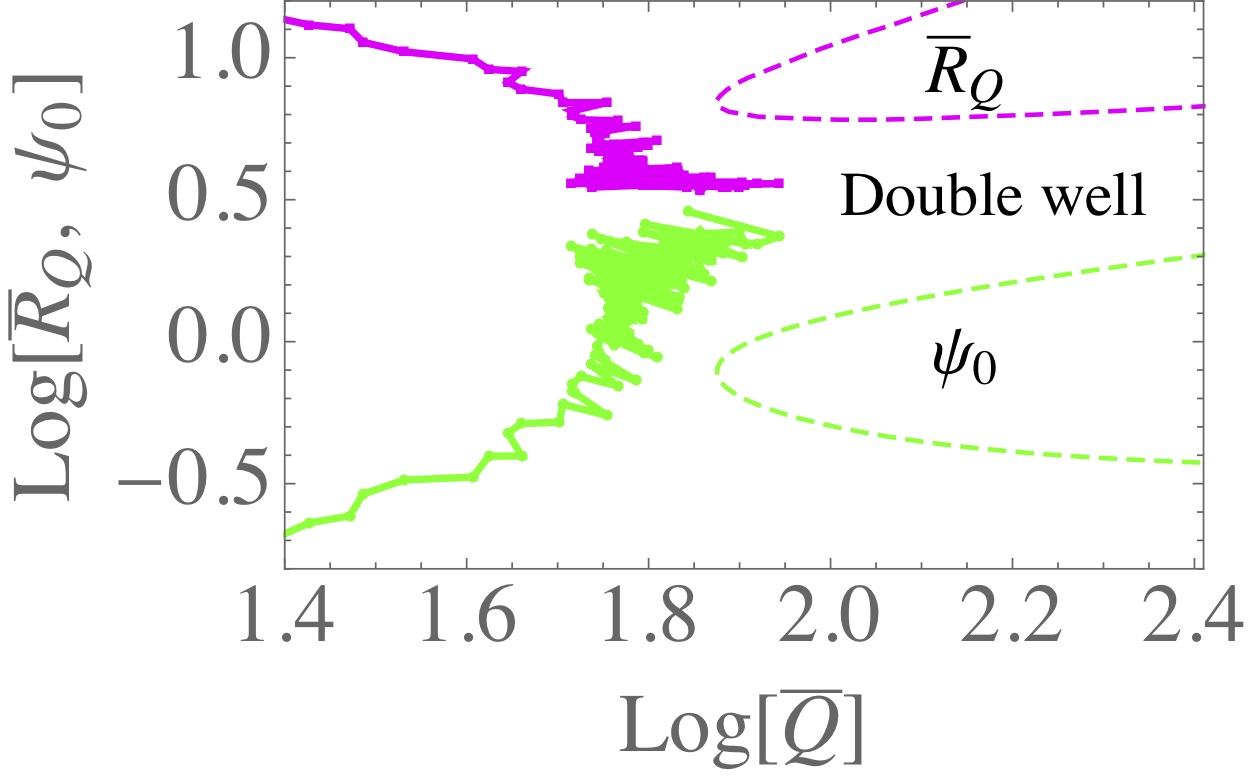}  
\caption{\small
Same as Fig.~\ref{fig:evolution1} but for a theory with the double well potential. 
We take $\phi_{\rm ini} = -1$ and $R_{\rm ini} = 7$. 
}
  \label{fig:evolution3}
\end{figure}

We take $\phi_{\rm ini} = -1$ and $R_{\rm ini} = 7$ for the initial condition. 
The resulting time-evolution of $\psi_0$, $\bar{R}_Q$, $E$, 
and $\bar{Q}$ are plotted in the left panel in Fig.~\ref{fig:evolution3}. 
The values of $\psi_0$ and $\bar{R}_Q$ are plotted as a function of $\bar{Q}$ in the right panel in Fig.~\ref{fig:evolution3}. 
We also plot theoretical results as dashed lines, where 
we include up to eighth order terms 
in the effective potential 
and the leading order correction of the difference between $m_\phi$ and $\omega$ ($= m_\phi - \mu$) (see App.~\ref{sec:higher order}). 
We take into account a leading order derivative correction 
up to mass dimension ten. 
The consistency with the theoretical result is not so good as the previous examples 
because 
higher-order terms in the effective potential may be relevant 
when $g_3 = \mathcal{O}(1)$. 
However, qualitative features are still consistent with the numerical simulation. 
In particular, 
I-ball/oscillon suddenly decay when 
$\bar{Q}$ decreases down to a certain critical value.

\section{Conclusions and Discussion}
\label{sec:conclusion}

In this paper, we have investigated the reason for the longevity of the I-balls/oscillons  in real scalar field theories, which is apparently mysterious because there are no evident conserved charges.  The point is that these localized configurations can be well described non-relativistically. As a result, an approximate symmetry emerges in that regime, which reflects the approximate number conservation, although the number is not conserved relativistically. The longevity of I-balls/oscillons can be understood as the smallness of the breaking of this approximated symmetry.

To clarify this viewpoint, we have explicitly constructed an effective theory for non-relativistic dynamics by integrating out relativistic modes. The resulting effective theory comes to have an approximate global U$(1)$ symmetry, whose charge is the number of particles. The breaking of U$(1)$ symmetry is encoded in the imaginary part of this effective action, which indicates the production of relativistic particles via number violating processes. We have shown that the profile of I-balls/oscillons can be obtained as far as the effect of the imaginary part can be safely neglected. Therefore, the profile of the I-balls/oscillons can be viewed as a projection of Q-balls onto the one axis (\textit{e.g.,} real part of a complex scalar field $\Phi_Q$) in the field space. We have demonstrated how to obtain a family of I-balls/oscillons as a function of the charge $Q$ for a given potential, and clearly shown that there exists a critical value of the charge, $Q_\text{cr}$, below which we do not have stable solutions.

Then, we discuss the \textit{classical} decay of I-balls/oscillons. As in the case of computing a decay rate of unstable particles, we can estimate the decay rate of I-balls/oscillons, by first regarding them as stable and then computing the imaginary part perturbatively. Since we are interested in the \textit{classical} decay, we omit all the loop contributions to the effective action. We have justified the assumption that the decay rate is much slower than the mass scale of the scalar field, by explicitly showing that the \textit{classical} decay rate is exponentially suppressed. Moreover,  we have first found that there are some critical values of charge (energy) at which the dominant decay channel vanishes accidentally. As a result, I-balls/oscillons stay at such critical points in most of their lifetime. In other words, I-balls/oscillons have approximate fixed points in their dynamics. We have ensured this essential nature of I-balls/oscillons by numerical simulations of the original relativistic scalar field theory. Our results indicate that the dynamical emergence of exponential separation of time scales leads to the approximately conserved quantity and the extraordinary longevity of the I-balls/oscillons.

While the main purpose of this paper is to clarify the \textit{classical} longevity of I-balls/oscillons, it is desirable to include \textit{quantum} effects if one would like to apply the I-balls/oscillons in cosmology. In our effective theory, the task is rather straightforward; that is, the inclusion of the loop diagrams. The real part of the loop contributions typically results in logarithmic corrections to the effective potential due to the renormalization running of coupling constants. The imaginary part of loop contributions has interesting effects. Since quantum fluctuations admit spontaneous relativistic particle production, the decay channels increase~\cite{Hertzberg:2010yz,Kawasaki:2013awa}. For instance, the production of relativistic modes with $p_0 \simeq 2 \omega$ and $p \simeq \sqrt{4 \omega^2 - m_\phi^2}$ can occur even in the $\mathbb{Z}_2$ symmetric case, which is otherwise prohibited as one can see from Fig.~\ref{fig:spectrum1}. The \textit{quantum} decay rate is suppressed by the smallness of $\epsilon \equiv (m_\phi - \omega)/m_\phi$~\cite{Hertzberg:2010yz,Mukaida:2014oza}, while there is no exponential suppression since we do not need the gradients to hit the pole contrary to the \textit{classical} decay [Eq.~\eqref{dQdt Gaussian}]. All one have to do is to include this term in the right-hand-side of Eq.~\eqref{dQdt Gaussian}. Then we can estimate the lifetime of I-balls/oscillons including \textit{quantum} effects.

Since we only assume that the configuration is dominated by the non-relativistic one, our formalism is rather generic. Therefore, it has a potential applicability to the formation of I-balls/oscillons, if one estimates the imaginary part of the effective action during the formation
and could guarantee its smallness. This is a clear contrast to the $\epsilon$ expansion where one expands the field around the I-balls/oscillons solutions. We come back to this issue elsewhere.
In addition, our formalism can also be used for rather generic condensates of non-relativistic scalar fields. For instance, it may be useful to study condensates of ultra light bosonic dark matter, such as fuzzy dark matter, axion stars, and so on~\cite{Ruffini:1969qy, Hogan:1988mp,Kolb:1993zz,Seidel:1993zk,Hu:2000ke,Guzman:2006yc,Sikivie:2009qn,Barranco:2010ib,Erken:2011dz, Chavanis:2011zm, Eby:2014fya, Berges:2014xea, Davidson:2014hfa, Guth:2014hsa, Eby:2015hyx, Braaten:2015eeu, Braaten:2016kzc, Hui:2016ltb, Eby:2016cnq}.

\section*{Acknowledgments}
\noindent
{\small 
We thank Masahiro Ibe and Wakutaka Nakano for pointing out typos and corrections in the figures.
Feynman diagrams were drawn by means of~\cite{Ellis:2016jkw}.
This work is supported by Grant-in-Aid for Scientific Research from the Ministry of Education,
Science, Sports, and Culture (MEXT), Japan, 
World Premier International Research Center Initiative (WPI Initiative), MEXT, Japan (K.M.),
and JSPS Research Fellowships for Young Scientists (K.M., M.T., and M.Y.).
}

\appendix
\section{Interaction terms}

Here we summarize all the interaction terms between NR modes and the other modes
for the sake of completeness.
First, let us consider the case of 
\begin{align}
	V_\text{int} = \frac{g_4}{4} \phi^4 + \frac{g_6}{6} \phi^6.
\end{align}
Note here that the sign of $g_4$ is opposite to that in the main text.

In this case, we have the following interaction terms.
For $ g_4$, we have
\begin{align}
	\mbox{\includegraphics[width=0.95\textwidth]{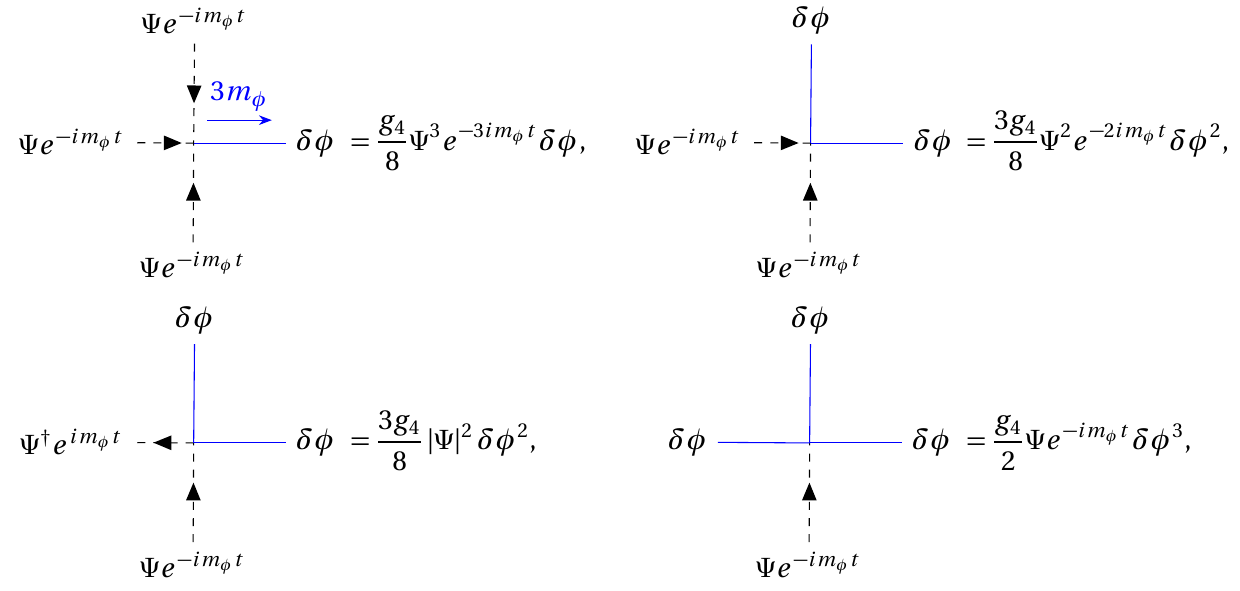}}
\end{align}
and those Hermite conjugates.
For $g_6$, we have
\begin{align}
	\mbox{\includegraphics[width=0.8\textwidth]{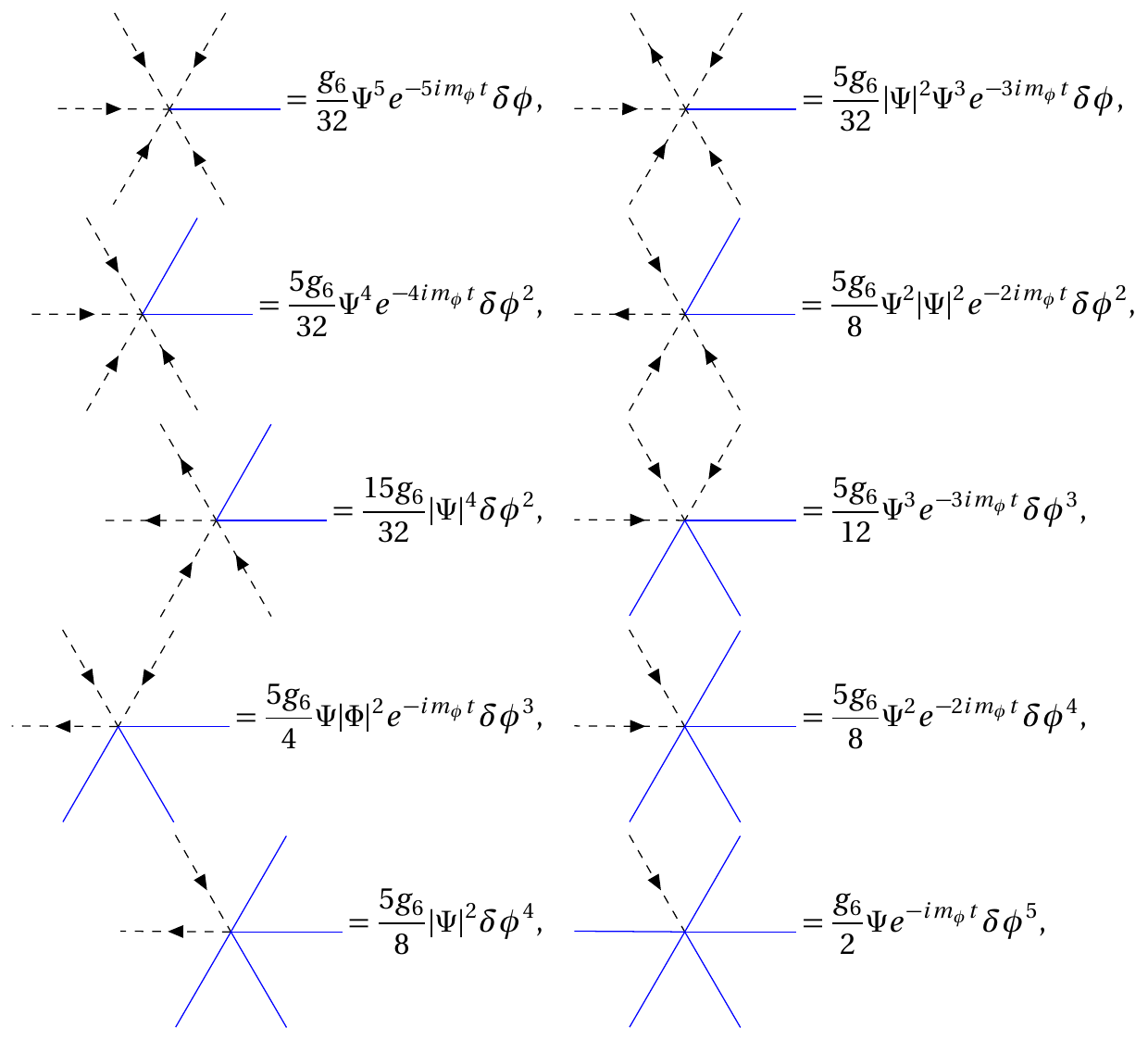}}
\end{align}
and those Hermite conjugates.
Here we have omitted $\Phi$ and $\delta \phi$ in the legs of feynman diagrams for brevity.

As an example of $\mathbb{Z}_2$ breaking interaction, let us consider 
\begin{align}
	V_\text{int} = \frac{g_3}{3} \phi^3.
\end{align}
The interaction terms are the followings:
\begin{align}
	\mbox{\includegraphics[width=0.62\textwidth]{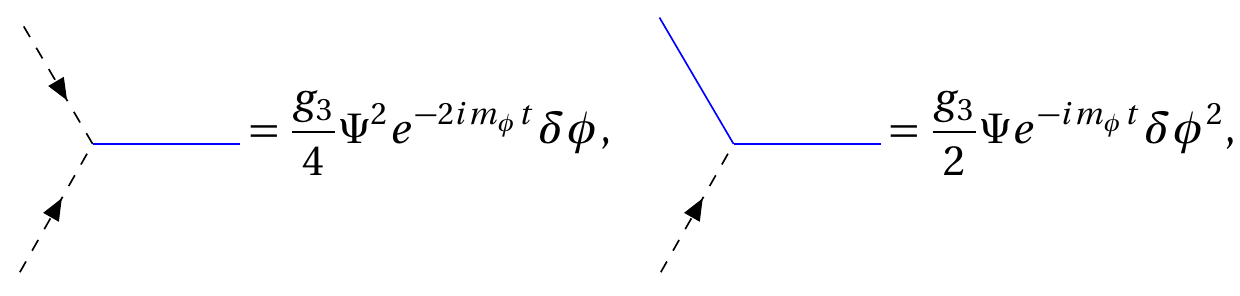}}
\end{align}
and those Hermite conjugates.
We also have terms involving $\delta \phi_V$ in the case of $\mathbb Z_2$ breaking term:
\begin{align}
	\mbox{\includegraphics[width=0.4\textwidth]{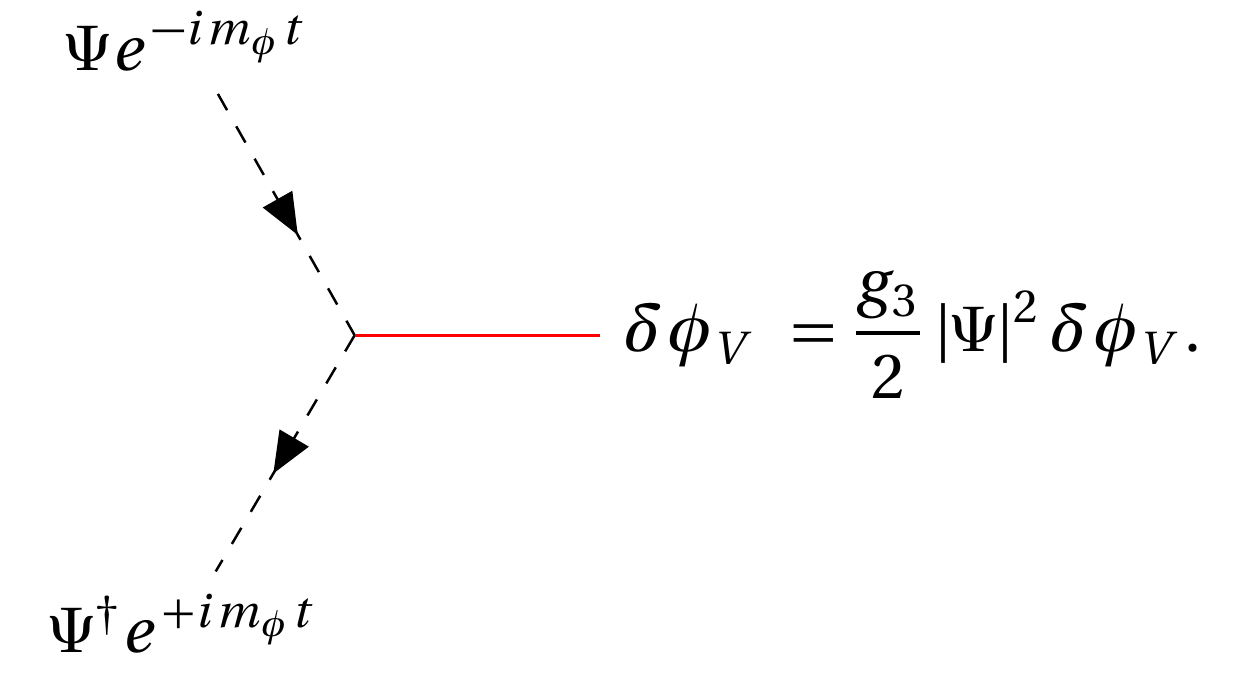}}
\end{align}

\section{Higher order terms}\label{sec:higher order}
Here we summarize higher order terms in the effective action.
Those terms are relevant if one applies our formalism to the double well potential.
This is because the non-relativistic condition is marginally satisfied in this case,
which calls for higher order terms in the non-relativistic expansion.

To be concrete,
we start with a relativistic scalar field theory with the following potential:
\begin{align}
	V_{\rm int} (\phi) = \frac{g_3}{3} \phi^3 + \frac{g_4}{4} \phi^4 
	.
\end{align}
Here, note that there is a mass term in addition to this potential in our notation [see \eq{lagrangian}]. 
By performing the procedure described in Sec.~\ref{sec:non_rela_eft},
we obtain the following effective potential for non-relativistic configurations:
\begin{align}
	V_\text{eff} (\abs{\Psi}) 
	= \sum_{n = 2} c_{2 n} \abs{ \Psi }^{2n}
	+ f (\abs{\Psi}, \nabla).
\end{align}
Note that the second term stands for corrections involving spatial gradients.

Up to the 8th dimensions, the coefficients are given by
\begin{align}
	c_4 &= 
	\frac{3g_4}{8}
	-\frac{5 g_3^2}{12 m_\phi^2}
	+\frac{2 (\mu/m_\phi)  g_3^2}{9 m_\phi^2}
	,
	\\
	c_6 &=
	\frac{g_4^2}{128 m_\phi^2}
	+\frac{9 (\mu/m_\phi)  g_4^2}{512 m_\phi^2}
	+\frac{53 g_4 g_3^2}{96 m_\phi^4} 
	-\frac{55g_3^4}{288 m_\phi^6}
	-\frac{421 (\mu/m_\phi)  g_3^4}{3456 m_\phi^6}
	,\\
	c_8 &= 
	-\frac{8245 g_3^6}{62208 m_\phi^{10}}
	-\frac{3647 (\mu/m_\phi)  g_3^6 }{124416 m_\phi^{10}}
	+\frac{3 g_4^3}{2048 m_\phi^4}
	+ \frac{27 (\mu/m_\phi)  g_4^3}{4096 m_\phi^4}
	\nonumber \\
	& \quad 
	-\frac{737 g_4^2g_3^2}{1024 m_\phi^6}
	+\frac{13877 (\mu/m_\phi)  g_3^2 g_4^2}{18432 m_\phi^6}
	+\frac{10775 g_4 g_3^4}{13824 m_\phi^8}
	+ \frac{3181 (\mu/m_\phi)  g_3^4 g_4}{82944 m_\phi^8}
	. 
\end{align}
Here, we include the leading order correction coming from the difference between 
the ``bare mass" $m_\phi$ and the ``effective mass" $\omega$ ($ = m_\phi - \mu$), 
which is much smaller than unity in most cases we are interested in. 
One can see that the expansion is good when $g_3 \ll 1$ 
because of small coefficients. 
The terms associated with the coupling $g_3$ and having $\mathcal{O}(1)$ coefficients 
come from the integration of non-relativistic mode $\delta \phi_V$ [see \eq{eq:2to1to2}]. 
Since $g_3 = \mathcal{O}(1)$ in the case of double well potential, 
the higher order terms are relevant and the expansion may not be good unless $\phi \ll 1$. 
Still, we expect that the qualitative features can be extracted from the effective potential 
even if we truncate it at some order. We use it up to 8th orders in this paper.

Since I-balls/oscillons are non-relativistic objects, the spacial derivative is usually smaller than 
the oscillation time scale. This allows us to neglect higher-order derivative terms. 
Here, we write the function $f$ 
associated with the derivatives up to the second-order derivatives and the mass dimension $8$: 
\begin{align}
	 \left. f (\abs{\Psi}, \nabla) \right\vert_{\rm mass\ dim \ 6}
	 &= 
	 \frac{g_3^2}{2 m_\phi^4} 
	 \Bigg[
	 (\nabla \abs{\Psi}^2 ) \cdot (\nabla  \abs{\Psi}^2 )  
	 +  \frac{1}{18} ( \nabla \Psi^\dag{}^2 ) \cdot ( \nabla \Psi^2 )
	 \Bigg]  
	 \\
	 \left. f (\abs{\Psi}, \nabla) \right\vert_{\rm mass\ dim \ 8}
	 &= 
	 \frac{g_3^4}{m_\phi^8}
	 \Bigg[
	 	\frac{1}{2304}(\nabla \Psi^\dag{}^3) \cdot (\nabla \Psi^3)
		+ \frac{7}{864} \prn{ \abs{\Psi}^2 \Psi^2 \Delta \Psi^\dag{}^2 + \abs{\Psi}^2 \Psi^\dag{}^2 \Delta \Psi^2} \nonumber \\
		&\qquad\qquad\qquad\qquad\qquad\qquad\qquad\qquad\qquad
		+ \frac{19}{36} (\nabla \abs{\Psi}^4) \cdot (\nabla \abs{\Psi}^2)
	 \Bigg] \nonumber\\
	 &+
	 \frac{g_3^2 g_4}{m_\phi^6}
	 \Bigg[
	 	\frac{1}{768} (\nabla \Psi^\dag{}^3) \cdot (\nabla \Psi^3) 
		+ \frac{5}{192} \prn{ \abs{\Psi}^2 \Psi^2 \Delta \Psi^\dag{}^2 +  \abs{\Psi}^2 \Psi^\dag{}^2 \Delta  \Psi^2 } \nonumber\\
		&\qquad\qquad\qquad\qquad\qquad\qquad\qquad\qquad\qquad
		- \frac{5}{4} (\nabla \abs{\Psi}^4) \cdot (\nabla \abs{\Psi}^2)
	 \Bigg] \nonumber\\
	 & +
	 \frac{g_4^2}{1024 m_\phi^4} (\nabla \Psi^\dag{}^3) \cdot (\nabla \Psi^3). 
\end{align}
We also write only relevant derivative terms with the mass dimension $10$, which come from the integration of $\delta \phi_V$: 
\begin{align}
	 \left. f (\abs{\Psi}, \nabla) \right\vert_{{\rm mass\ dim \ 10 \ from \ }\delta \phi_V}
	 &= 
	 \frac{g_3^6}{m_\phi^{10}}
	 \Bigg[
	 	\frac{3539}{6912} (\nabla \abs{\Psi}^6) \cdot (\nabla \abs{\Psi}^2)
		+
	 	\frac{361}{2592} (\nabla \abs{\Psi}^4) \cdot (\nabla \abs{\Psi}^4)
	 \Bigg] \nonumber\\
	 &+
	 \frac{g_3^4 g_4}{m_\phi^8}
	 \Bigg[
	 	-\frac{1879}{768} (\nabla \abs{\Psi}^6) \cdot (\nabla \abs{\Psi}^2)
		-
	 	\frac{95}{144} (\nabla \abs{\Psi}^4) \cdot (\nabla \abs{\Psi}^4)
	 \Bigg] \nonumber\\
	 &+
	 \frac{g_3^2 g_4^2}{m_\phi^6}
	 \Bigg[
	 	\frac{1697}{1024} (\nabla \abs{\Psi}^6) \cdot (\nabla \abs{\Psi}^2)
		+
	 	\frac{25}{32} (\nabla \abs{\Psi}^4) \cdot (\nabla \abs{\Psi}^4)
	 \Bigg].
\end{align}
\small
\bibliography{ball}

\end{document}